\documentclass[review]{elsarticle}

\usepackage{lineno,hyperref}
\usepackage{amsmath}            
\usepackage{lineno,hyperref}
\usepackage{subfig}
\usepackage{multirow} 
\usepackage{multicol} 
\usepackage{array}
\usepackage{pdflscape}
\usepackage{xcolor}
\usepackage[flushleft]{threeparttable} 
\usepackage{booktabs}
\usepackage{xfrac}
\usepackage{cleveref}
\usepackage{graphicx}
\usepackage{color,soul}
\soulregister\cite7
\soulregister\ref7
\soulregister\eqref7
\soulregister\pageref7

\graphicspath{{images/}}

\newcommand{\Fref}[1]{Figure~\ref{#1}}

\journal{Acta Materialia}

\begin{document}

\begin{frontmatter}

Authors’ Accepted Manuscript

\bigskip
\bigskip
\bigskip
\bigskip
\bigskip
\bigskip
\bigskip

\leftskip=0cm plus 0.5fil \rightskip=0cm plus -0.5fil
\parfillskip=0cm plus 1fil \large \textmd{\textbf{Grain Boundary Serration in Nickel Alloy Inconel 600: Quantification and Mechanisms
}} 

\bigskip
\bigskip
\normalsize
Yuanbo T. Tang$^{a}$, Phani Karamched$^{a}$, Junliang Liu$^{a}$, Jack C. Haley$^{a}$, 
Roger C. Reed$^{a,b}$, Angus J. Wilkinson$^{a}$

\bigskip
$^{a}$Department of Materials, University of Oxford, Parks Road, Oxford, OX1 3PH, United Kingdom

$^{b}$Department of Engineering Science, University of Oxford, Parks Road, Oxford, OX1 3PJ, United Kingdom

\bigskip
\bigskip
\bigskip
\bigskip
\bigskip
\bigskip
The published version of record can be found in 

Acta Materialia, (2019) \textbf{181}, 352-366    https://doi.org/10.1016/j.actamat.2019.09.037

\bigskip
\bigskip
\bigskip
\bigskip
\bigskip
\bigskip

This version of the article is shared here under the Creative Commons CC-BY-NC-ND license
(https://creativecommons.org/licenses/by-nc-nd/4.0/)

\newpage

\title{Grain Boundary Serration in Nickel Alloy Inconel 600: Quantification and Mechanisms}


\author[addressmain]{Yuanbo T. Tang}
\author[addressmain]{Phani Karamched}
\author[addressmain]{Junliang Liu}
\author[addressmain]{Jack C. Haley}
\author[addressmain,addresstwo]{Roger C. Reed}
\author[addressmain]{Angus J. Wilkinson}

\address[addressmain]{Department of Materials, University of Oxford, Parks Road, Oxford, OX1 3PH, United Kingdom}
\address[addresstwo]{Department of Engineering Science, University of Oxford, Parks Road, Oxford, OX1 3PJ, United Kingdom}

\begin{abstract}

The serration of grain boundaries in Inconel 600 caused by heat treatment is studied systematically. A new method based on Fourier transforms is used to analyse the multiple wave-like character of the serrated grain boundaries. A new metric -- the serration index -- is devised and utilised to quantify the degree of serration and more generally to distinguish objectively between serrated and non-serrated boundaries. By considering the variation of the serration index with processing parameters, a causal relationship between degree of serration and solution treatment/cooling rate is elucidated. Processing maps for the degree of serration are presented.
Two distinct formation mechanisms arise which rely upon grain boundary interaction with carbides: (i) Zener-type dragging which
hinders grain boundary migration and (ii) a faceted carbide growth-induced serration.

\end{abstract}

\begin{keyword}

\texttt Grain Boundary Serration \sep HR-TKD \sep HR-TEM \sep Fourier Transform \sep Heat Treatment \sep Carbide \sep Processing Map

\end{keyword}

\end{frontmatter}


\section{Introduction}

A curiosity of the nickel-based superalloys -- in common with other alloy systems -- is that their grain boundaries can become serrated \cite{Pickering2012, Li2015, Carter2015, Alabort2018}. When this happens, the deformation behaviour can be altered \cite{Telesman2004, Kontis2017}, which offers the opportunity for so-called grain boundary engineering. But unfortunately, the precise details of the physical mechanisms which cause the serration effect are not well understood at present. From a practical perspective, it is acknowledged that serration can be triggered by heat treatment, although precise quantitative information is lacking. Moreover, although the promise of improving properties by grain boundary engineering in this way is obviously attractive, little progress has been made so far. Furthermore, it is not clear how potent the effect on behaviour might be.

What is needed is a much improved understanding of the underlying mechanism. Koul \& Gessinger \cite{Koul1983} -- by making observations on a range of alloys -- were the first to identify the importance of the $\gamma'$ phase and the net strain energy difference across two boundaries. Their ideas were modified by Henry \textit{et al} \cite{Henry1993} to incorporate solute transport, which was deemed to assist preferential growth of coherent $\gamma'$, for which later work by Mitchell \textit{et al} \cite{Mitchell2009} provided further validation. On the other hand -- particularly for low $\gamma'$ fraction alloys -- alloying
with Cr and C has also been found to be influential, presumably because of the precipitation of carbide phases \cite{Jiang2012, Lim2014}. Based on these observations, other mechanisms on oriented growth and discontinuous precipitation have also been proposed. Hence at this stage, most explanations for how grain boundaries become distorted into irregular morphologies invoke the precipitation of intergranular precipitates, such as $\gamma'$, $\rm M_{23}C_6$ carbide, $\delta$, etc \cite{Koul1983, Henry1993, Jiang2012, Hong2009, Hong2012, Yeh2011}. But the details of the heat treatment also matter. Several researchers report success in generating serrated grain boundaries using supersolvus solution treatment. Second phases re-precipitating along grain boundaries have been associated with the initiation of serration, and the importance of coherency relationships between the decorating particle and the matrix have been emphasised \cite{Mitchell2009, Jiang2012, Hong2009, Yeh2011, Hong20122}.

A difficulty with all work thus far is that the methods used to assess the degree of serration are usually subjective; neither a definition for serration nor a criterion to assess its severity is available, which is obviously unsatisfactory.
Processing parameters have thus been only vaguely and qualitatively correlated to microstructure as the key factors influencing serration are not well isolated. In this paper, this difficulty is circumvented by proposing a new quantification method which makes use of Fourier transformation of the grain boundary traces, which will be shown to improve the objectivity with which serration can be assessed; this approach allows an index to describe the degree of serration to be proposed. This development -- along with
experimentation on the alloy IN600 with many different heat treatments used to develop varying degrees of serration -- has allowed fundamental insights into the mechanism of serration to be revealed.

\section{Experimental Methods}

The nickel-based alloy Inconel 600, see Table~\ref{table: composition}, was chosen for study, on the basis of its relatively uncomplicated microstructual phases -- namely $\gamma$ matrix with Cr-rich carbides and the absence of the $\gamma'$-forming elements.
In this case therefore, the well known $\gamma'$ phase is absent; as will be seen, this makes the findings easier
to interpret so that firm conclusions can be more easily drawn.

\subsection{Grain Boundary Serration via Heat Treatment}

\begin{landscape}
\begin{figure}
  \centering
  \includegraphics[width=1.5\textwidth]{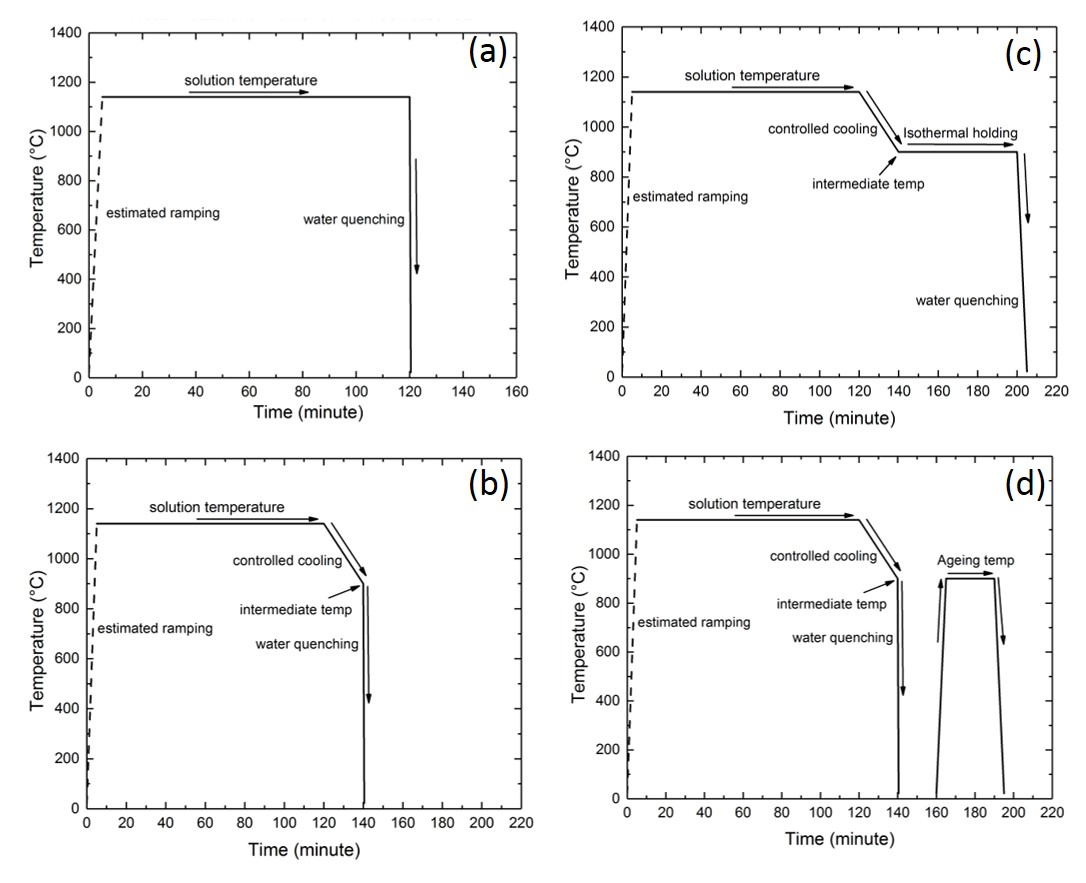}
  \caption{Four types of heat treatment applied to the material. (a) Solution treatment followed by constant cooling rate (air/water quenching), (b) solution treatment followed by slow cooling to an intermediate temperature for quenching, (c) solution treatment followed by slow cooling to an intermediate temperature and isothermal holding before quenching and (d) additional ageing added to (b).}
  \label{fig:HT}
\end{figure}

\end{landscape}

Classical heat treatment studies have been used to induce different degrees of serration, by varying solution heat treatment temperature, 
cooling rate and the details of an isothermal step. Heat treatments were carried out using a Carbolite TZF 15/610 tube furnace, 
with temperatures and ramping rates calibrated by N-type thermocouples. \Fref{fig:HT} illustrates the four distinct types of
heat treatment which have been used in this study, which have been labelled A, B, C and D. A summary of related data is given
in Table~\ref{HT_matrix}. Heat treatments labelled A involved quenching from above the carbide solution temperature and then 
cooling rapidly; these were used to determine the influence of carbide solutioning. Heat treatments labelled B (the majority) involved cooling at a constant rate of between 0.25 and 60 deg C/min from a temperature between 1000 and $1140^\circ$C to an intermediate temperature -- most typically $900^\circ$C -- before water quenching immediately. Heat treatments C followed those of B with the exception that water quenching occurred after a hold period at the intermediate temperature. Finally, heat treatment D involved reheating to an ageing temperature of $900^\circ$C. Then first column of Table~\ref{HT_matrix} contains the
label for each heat treatment which will be referred to subsequently -- for example, B1, B2, B4 and B9 refer to a set of heat treatments involving solutioning at 1000, 1070, 1100 and $1140^\circ$C respectively followed by a common cooling rate of 
$12^\circ$C/min to $900^\circ$C. 

\subsection{Scanning electron microscopy based characterisation and image processing}

A Zeiss Merlin field emission gun scanning electron microscope (FEG-SEM) was used for imaging studies. Back-scattered electron (BSE) images were collected with a step size of 46.5 nm at all conditions. For each microstructure, 15 randomly picked grain boundaries were captured for analysis. The grain boundary length that contributed for analysis was in excess of 1 mm for each specimen. Where serration were identified, images of intergranular carbides were taken for analysis aided by image processing, making use of compositional contrast of the BSE image to distinguish carbides from the matrix. ImageJ and Paint.net softwares were utilised for segmentation of the shape based on image intensity. Feret diameters were used as shape descriptors, where Feret max/min represents the maximum/miminum distance between two parallel planes touching both sides of the particle profile. Hence the Feret max and Feret min are quoted as carbide length and thickness respectively. Circularity is defined here as 4$\pi$ $\times$ area/perimeter$^2$, which is used to describe the aspect ratio of a carbide that indicates the stage of growth. Carbide analysis results from a range of the heat treatment cases are given in Table~\ref{carbide}.

\subsection{Transmission Electron Microscopy (TEM): Diffraction and Composition Profiling}

Foil samples of 3 mm diameter were prepared from as-received material and from B7 (serrated) material, and then thinned 
(to less than 100 nm) using a twin-jet electropolishing unit. Electropolishing was conducted using an electrolyte of 5$\%$ perchloric acid (HClO$_{4}$) in methanol at -4$0^{\circ}$C. Micrographs were captured using a JEOL JEM-3000F FEG-TEM microscope operated at 300 kV. To determine the crystal structure of the carbides, selected-area diffraction (SAD) patterns were recorded from each carbide on three different zone-axes. The patterns were indexed by measuring the spacing, d1 and d2, of two different g-reflections, and the angle between them. The appropriate hkl reflections were identified as the allowed reflections required to produce the correct angle and d1/d2 ratio. Complementary to the diffraction patterns, high-resolution TEM micrographs were also captured down the zone axis to facilitate interpretation of the patterns. Moreover, compositional information across the interface of matrix and carbide was examined using energy dispersive X-ray spectroscopy (EDS) under the scanning (STEM) mode with a step size of 1 nm.

\subsection{Transmission Kikuchi Diffraction (TKD) Experimentation}

\begin{landscape}
\begin{figure}
  \centering
  \includegraphics[width=1.2\textwidth]{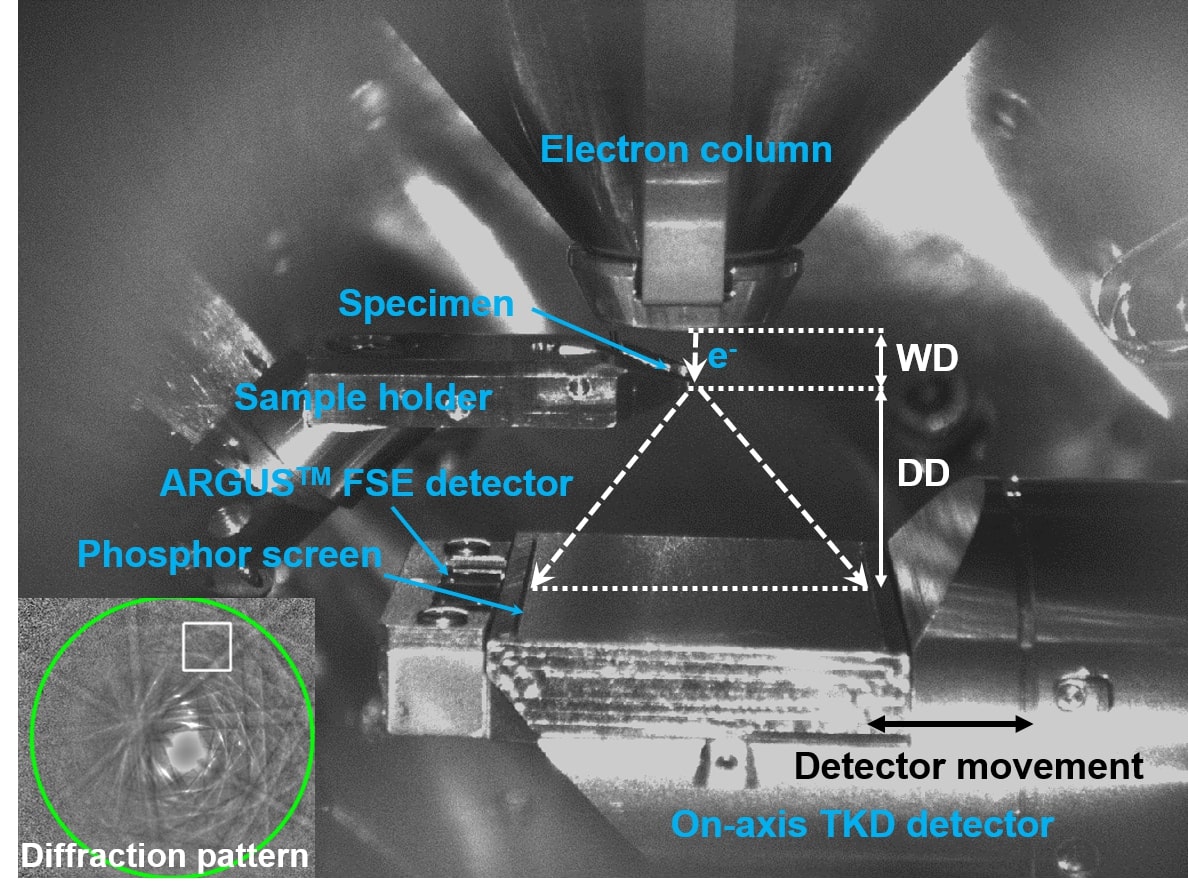}
  \caption{The Transmission Kikuchi Diffraction (TKD) setup for the current study with a typical EBSD diffraction pattern acquired using the technique.}
  \label{fig:TKD_setup}
\end{figure}
\end{landscape}

An overview of the on-axis SEM-TKD setup is displayed in \Fref{fig:TKD_setup}. A Zeiss Merlin FEG-SEM system was operated using 30 kV voltage with a probe current of 2 nA for the prepared foil samples. The working distance used was 5 mm and the step size ranged from 1 - 10 nm. An OPTIMUS$^{TM}$ TKD detector was placed beneath the sample. The integrated detecting system includes an ARGUS$^{TM}$ forescattered electron (FSE) detector and a horizontal phosphor screen. The ARGUS$^{TM}$ detector consisted of three Si diodes, enabling bright field imaging (central diode) and dark field imaging (side diodes) to provide strong diffraction and phase contrast. The detector can be moved in the horizontal direction, as illustrated in \Fref{fig:TKD_setup}, to align the beam and detector in use. The phosphor screen was used to record Kikuchi diffraction bands, whose pattern center was projected to the middle of the screen to minimise any pattern distortion. The pattern resolution used was 800 $\times$ 600 in 16 bits, which facilitated further lattice strain and rotation analysis using a cross-correlation based technique \cite{Wilkinson2006, Wilkinson2010, Britton2012}. The cross-correlation analysis was conducted at 19 regions of interest (ROI) distributed in a circle to avoid the central part of the on-axis TKD patterns and the bright spot features captured there. The cross-correlation analysis generates maps of the lattice strain and lattice rotation variations relative to a reference point selected within each grain. The lattice rotation gradients were used to estimate the geometrically necessary dislocation (GND) density using methods described in \cite{Wilkinson2010, Karamched2011}. To avoid any potential bending of the thin foils that might have undermined the accuracy of results, we investigated a relatively thick region which was about 15 $\mu$m away from the electro-polished hole.

\subsection{Quantification of Serration using Fast Fourier Transform (FFT) Analysis}

\begin{landscape}
\begin{figure}
  \centering
  \includegraphics[width=1.5\textwidth]{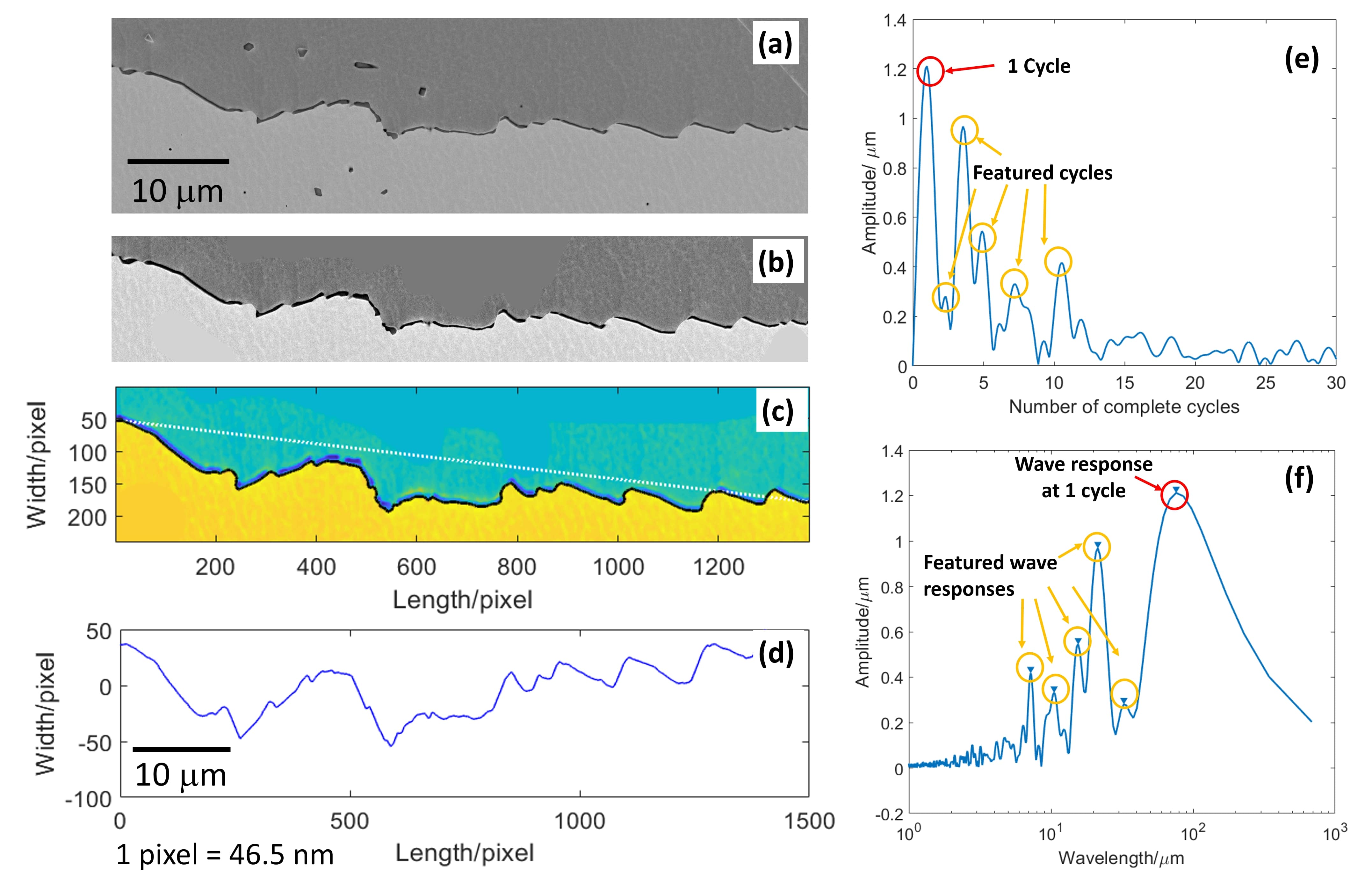}
  \caption{FFT analysis on grain boundary serration. (a) is the original image of a grain boundary and (b) is the digitally processed image using brightness/contrast and removal of artifacts. (c) is the binarised image with its boundary trace outlined and (d) is the baseline-corrected wave function. (e) is the Power/Frequency spectrum after transformation and (f) is spectrum in wavelength and amplitude.}
  \label{fig:FFT}
\end{figure}
\end{landscape}

\Fref{fig:FFT} illustrates step by step the fast Fourier Transform (FFT) \cite{Ametani1973, Tsinopoulos1999, Mitchell2009FFT} based method used here to quantify the degree of serration: (a) shows a typical captured grain boundary, here decorated by some precipitates. Image processing was performed to increase contrast and to remove interior precipitates/artifacts. The updated image (b) was then binarised using a threshold value in the histogram. Correction of the skewness is performed in (c), where a straight line is drawn connecting the end-to-end distance of the segmented curve representing the grain boundary. \Fref{fig:FFT} (d) illustrates the perpendicular distance between each pixel on the curve from the straight line between the start and the end points chosen. The curve displayed in (d) is then analysed using the FFT method, to enable frequencies of sinusoidal waves with corresponding amplitude to be extracted. Physical interpretation of the frequency in this case is the number of complete cycles that it exhibits in the image, as shown in \Fref{fig:FFT} (e). Lastly, with known distance of signal length and number of complete cycles, the wavelength of each wave present is obtained, see (f).

Each serrated grain boundary profile was transformed into its constituent sinusoidal waves. For interpretation, waves with an amplitude greater than 0.2 $\mu$m were carried further into the analysis. Additionally, the peak at 1 cycle per frame length was neglected for further analysis, as it represented smooth long-range undulation of the boundary. On average, each boundary profile was shown to possess of the order of seven intrinsic waves, and hence the method has demonstrated an improved objectivity of the boundary profile compared to approximating them as a single sinusoidal curve as has been done previously \cite{Mitchell2009, Hong2009}. However, a new metric is needed to describe the degree of serration, preferably combining the multiple wavelengths and amplitudes into a single scalar value. Because the serration amplitude increases with wavelength, as also observed elsewhere \cite{Hong2009}, the best way to reflect the significance of serration is by the area carved out relative to a smooth boundary. Therefore, the serration index relates to the area that the curve subtended, and can be thought of as follows. Three waves with the highest amplitudes are chosen to represent the overall shape of the serrated grain boundary. For each constituent wave, its wavelength and amplitude are multiplied -- three indices are obtained from each serration profile. The overall serration index of a particular profile is the average value of the three indices, in units of $\mu$m$^{2}$.

\section{Results}

\subsection{The Generation of Serrated Grain Boundaries}

\begin{landscape}
\begin{figure}
  \centering
  \includegraphics[width=1.5\textwidth]{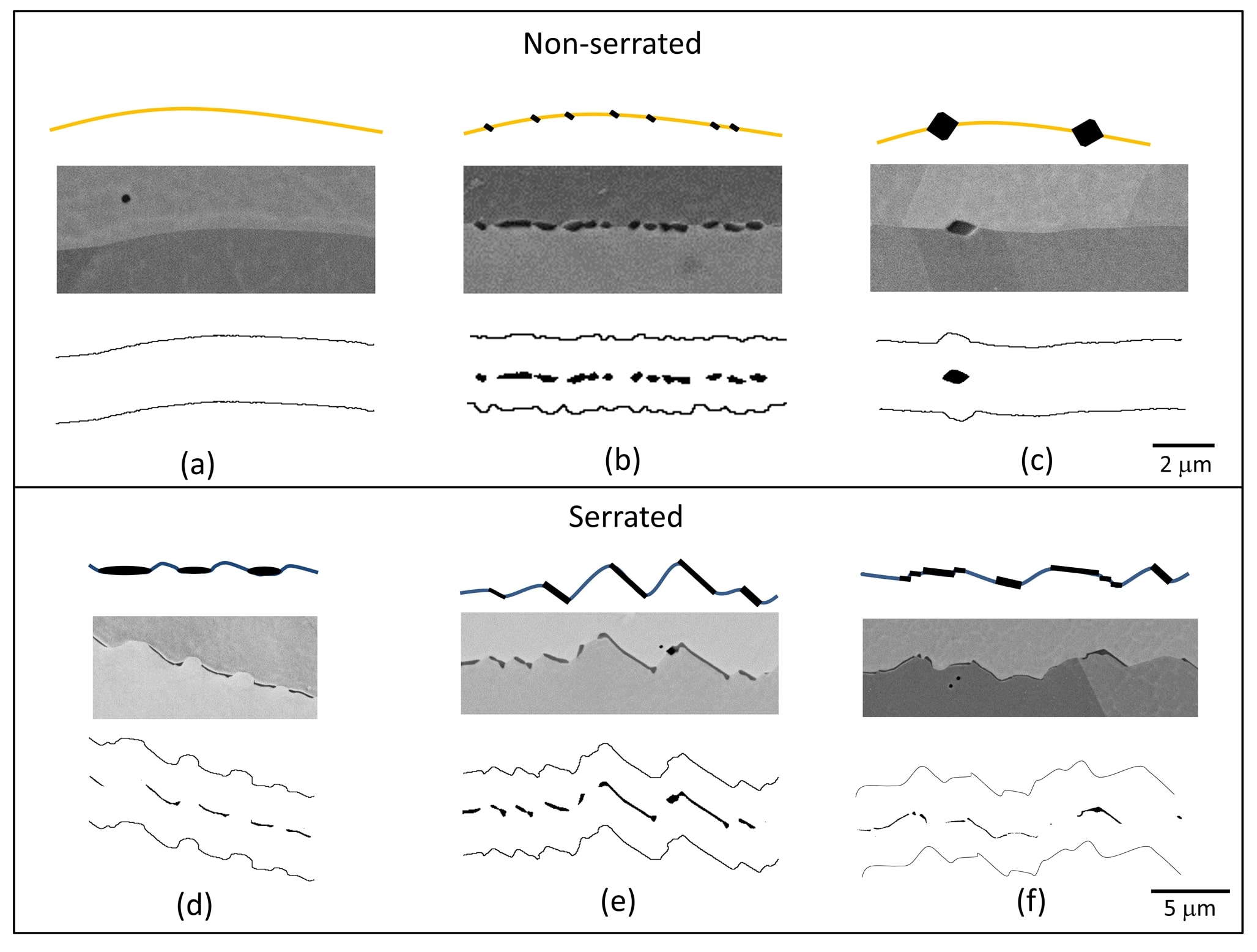}
  \caption{Summary of the six types of grain boundaries observed. A combination of schematic diagrams, representative SEMs and traces of boundary on the either side of grain boundaries illustrated this. (a-c) represent non-serrated grain boundary and (d-f) displayed serrated grain boundary. The specific heat treatment that yielded the boundaries presented in (a-f) are as-received, D1, as-received, B9, C2 and B4 respectively.}
  \label{fig:gb_sch}
\end{figure}
\end{landscape}

The approaches and techniques described above have generated -- as judged by imaging on the scale of scanning electron
microscopy -- grain boundaries which are either clearly very serrated, or else not serrated at all, see \Fref{fig:gb_sch}. Many hundreds of images have been analysed in order to come to the conclusion that several different categories of boundary arise from these heat treatment as summarised in \Fref{fig:gb_sch}. In this diagram, the images below each micrograph are
representations of the grain boundaries and also the locations of carbide phases on the boundaries where these exist. One can see that carbides are sometimes present on the boundaries but not always; moreover, the presence of carbides on boundaries is not
sufficient to cause serration. Additionally, we have noticed that where serration occurs, it is always associated with
random high angle grain boundaries. Twin boundaries ($\Sigma$3) appears to be immune from serration formation. The final column in Table \ref{HT_matrix} lists the type of grain boundaries (one of a, b, c, d, e or f of \Fref{fig:gb_sch}) generated by each specific heat treatment.

Thus, six types of grain boundaries have been identified. Of the six types, only (a) was completely clear of carbides so that can be unambiguously described as being non-serrated. The boundaries in (b) and (c) of \Fref{fig:gb_sch} contain intergranular carbides that grow on the grain boundary, but their smoothness is perturbed by irregularly shaped particles which may leave the false impression of serration. In contrast, for cases d to f of \Fref{fig:gb_sch}, all boundaries are clearly serrated with carbides obtaining coincident profiles with boundaries. In these cases, closer attention can then be paid to the boundary traces on both sides of the particle, as shown next to the actual boundary in \Fref{fig:gb_sch}. It can be readily seen by considering the difference between (b) $\&$ (c) vs (e - f) that the boundary traces are almost perfect mirror images of each other in the former cases. However, the boundaries have almost identical shape to each other in the latter cases. Here we propose a criterion for the determination of intrinsic serrations: the profile of boundaries along each side of particles should share the same trend of curvature and approximately match up once overlapped. On this basis, \Fref{fig:gb_sch} (b) $\&$ (c) are classified as non-serrated grain boundaries.

\begin{landscape}
\begin{figure}
  \centering
  \includegraphics[width=1.5\textwidth]{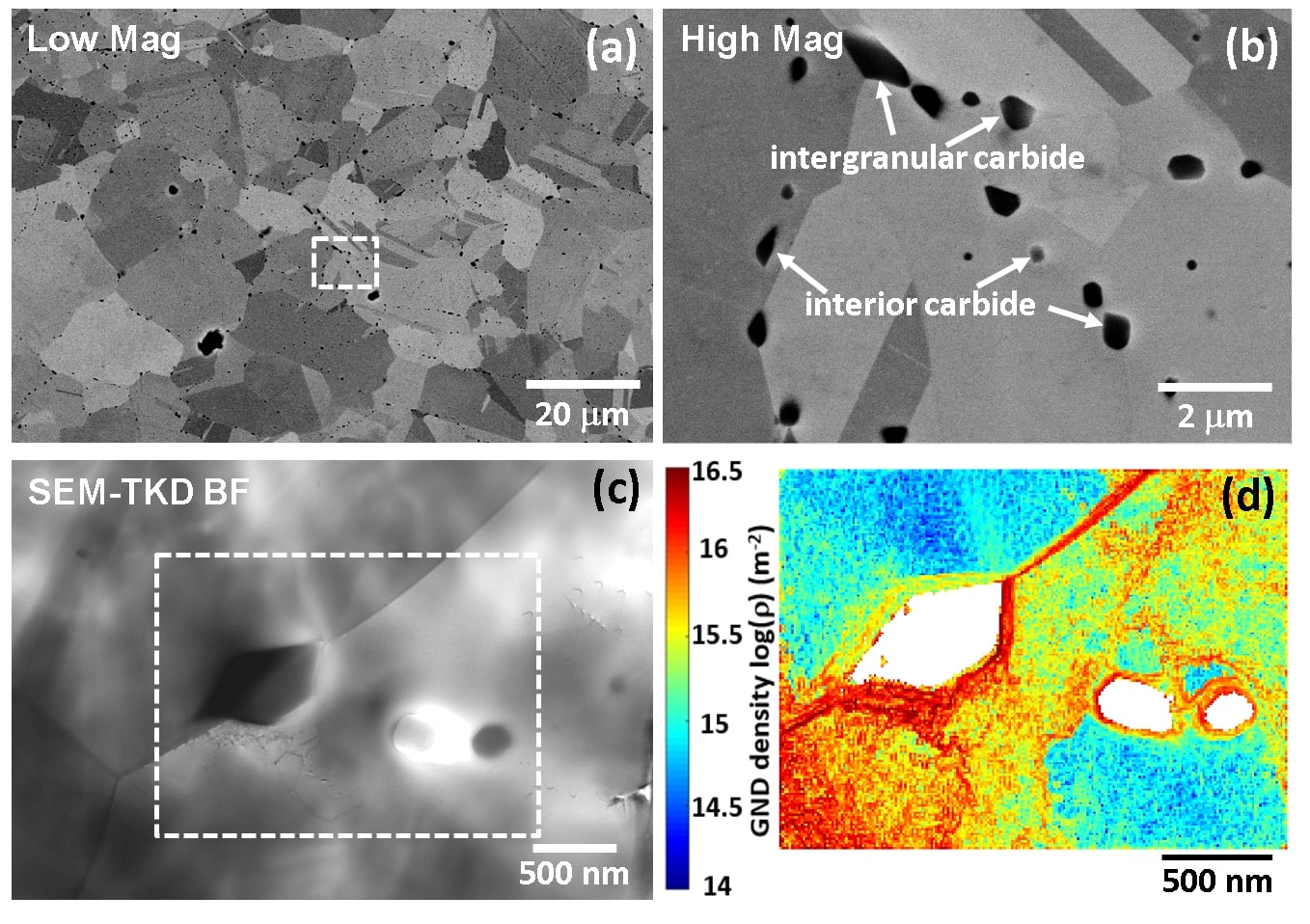}
  \caption{The microstructure overview of the as-received Inconel 600. (a) $\&$ (b): SEM BSE images obtained from two magnifications, where granular carbides were observed both on the grain boundaries and inside the grains. (c): TKD bright field image of an intergranular carbide and (d) represents the GND density map of the same region.}
  \label{fig:AR_micro}
\end{figure}

\end{landscape}

For the avoidance of doubt, \Fref{fig:AR_micro} shows the material in the as-received state. Two magnifications of SEM BSE images are illustrated in (a) $\&$ (b). Granular carbides decorate grain boundaries but are also present in grain interiors. TKD bright field imaging also revealed carbides along a grain boundary and in the matrix next to it, see \Fref{fig:AR_micro} (c), where a group of dislocations are evident between them. The geometrically necessary dislocation (GND) density map shown in (d) has revealed the expected strong correlation between the high GND density region and the location of dislocation clusters, seen in the bright field image. In what
follows below, it will be confirmed that the carbide phase is Cr-rich $\rm M_7 C_3$.

\subsection{Variation of The Degree of Serration With Heat Treatment Parameters}
\label{FFT_method}

Those heat treatment conditions of Table \ref{HT_matrix} which caused serration were studied in much greater detail in order
to extract quantitative information. Table \ref{carbide} summarises the serration indices measured for the serrated grain boundaries. Also given are a great deal of stereological data for the carbides associated with each boundary. A number of conclusions emerge.

\begin{figure}
  \centering
  \includegraphics[width=1.2\textwidth]{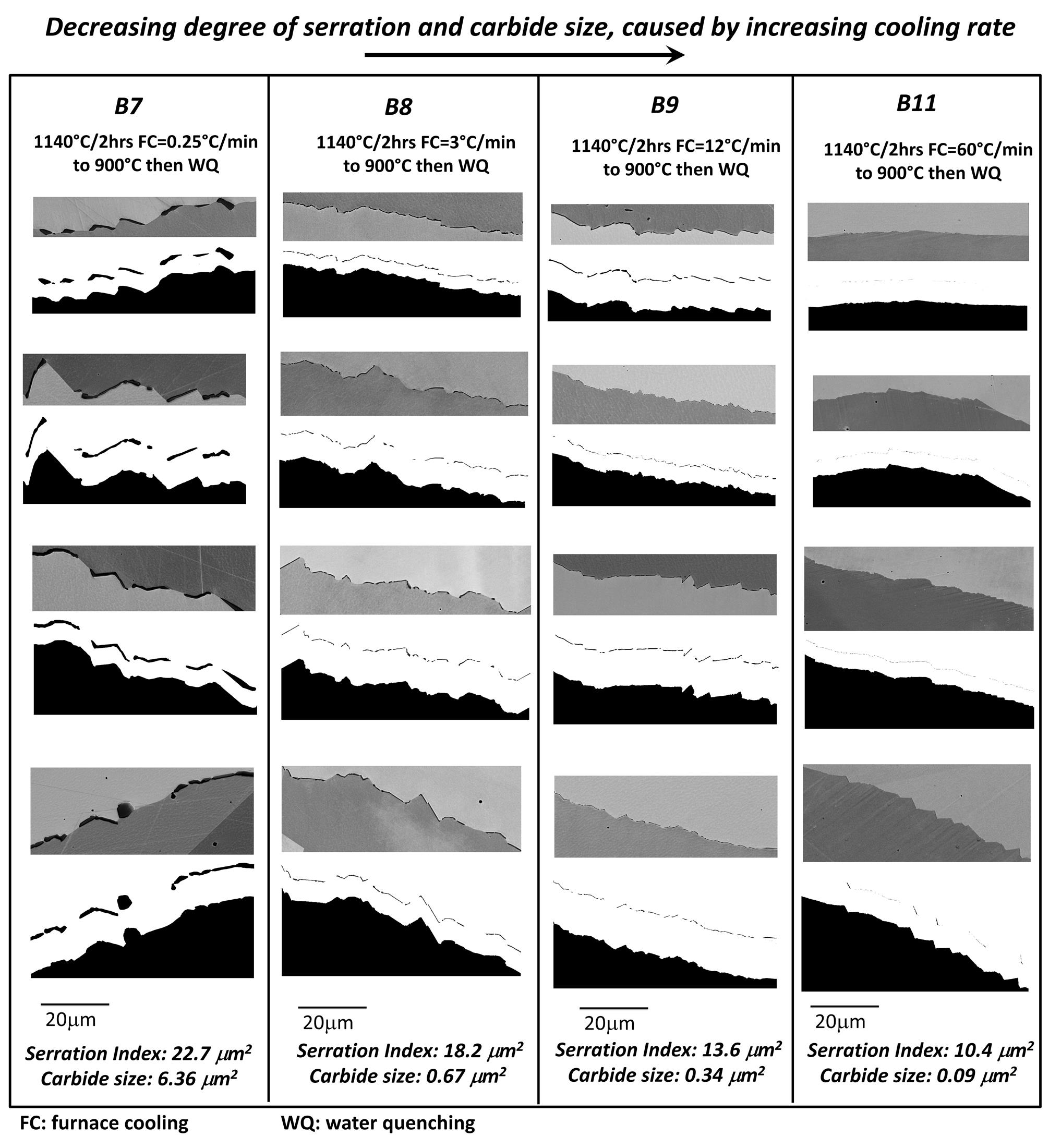}
  \caption{Typical grain boundaries studied in four heat treatment conditions that obtain serrated morphologies, \textit{i.e.} B7, B8, B9 $\&$ B11. For each boundary that displayed, intergranular carbide and boundary traces were outlined. Quantification of serration index and carbide size is also presented.}
  \label{fig:gb_prof}
\end{figure}

First, as confirmed by the data for the B7, B8, B9 and B11 conditions for which images are also given in \Fref{fig:gb_prof} ---
this is a set of experiments for which the solutioning temperature was maintained constant at $1140^\circ$C --- the serration index decreased from 22.65 to 10.35 $\mu$m$^{2}$ with increasing cooling rate. The carbide size also decreased substantially from 6.36 to 0.09 $\mu$m$^2$, with carbide length and thickness following similar behaviour. The carbide nucleation density was also affected strongly and increased from 109.2 to 696.6 mm$^{-1}$. Moreover, the circularity of carbide showed positive correlation with cooling rate, and slower cooling rates conferred larger carbides. The growth of carbide would seem to have a preferred orientation as it loses circularity as growth proceeds. Furthermore, the above observations on the serration indices and carbide numbers and morphologies are unaltered at the lower solution temperature of 1100$^\circ$C (B4 $\&$ B5). Not surprisingly, to a certain extent, high cooling rate will not be able to generate serration at all. The critical cooling rate for serration is associated with air cooling (measured at $600^\circ$C/min by N-type thermocouple); a cooling rate higher than this suppresses serration (heat treatment A1).

\begin{landscape}
\begin{figure}
  \centering
  \includegraphics[width=1.5\textwidth]{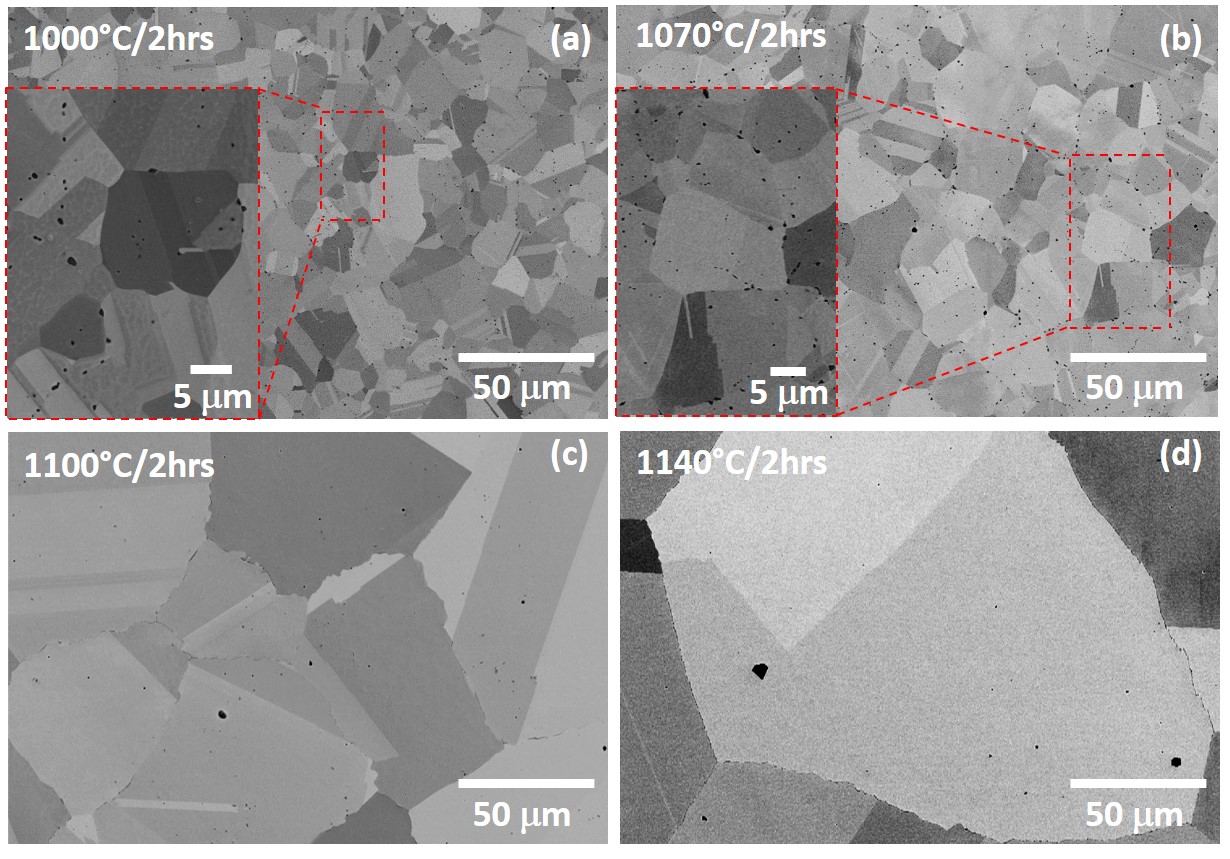}
  \caption{Effect of solution temperature on microstructure ranging from 1000 - 1140$^\circ$C. (a-d) are microstructure obtained from B1, B2, B4 and B9 respectively.}
  \label{fig:solution}
\end{figure}
\end{landscape}

Second, one can consider the effect of varying the solution heat treatment from $1000$ to $1140^\circ$C at fixed duration of 120 minutes. \Fref{fig:solution} illustrates the substantial influence of solution temperature on promoting grain growth. The results enabled the identification of the solution window that triggers serration. Limited carbide dissolution and grain growth was observed at $1070^\circ$C/120 mins (B2). However, notable grain growth and carbide dissolution were observed at $1100^\circ$C/120 mins (B4), accompanied with some degree of grain boundary serration. At constant cooling rate, the serration index and carbide size demonstrated positive correlation with solution temperature (B6 $\&$ B8 and B4 $\&$ B9). In addition, an isothermal hold confers some enhancement in generating both serration and carbide size (measured in area). For example, an isothermal hold at 900$^\circ$C for 30 mins increased the serration index from 13.62 to 16.15 $\mu$m$^2$ and the mean carbide size from 0.34 to 0.44 $\mu$m$^2$ (B9 $\&$ C2).

Finally, isothermal holding with subsequent ageing was found to engender grain boundary carbides, as shown in \Fref{fig:gb_sch} (b), but is not sufficient to yield serration since the carbides which then form are too fine (D1). By comparing C1 $\&$ C3 --- both conditions obtained the same solution treatment and intermediate temperature --- one can see that a slower cooling rate indicates that it is possible to yield serrated boundaries in association with intergranular carbides. This suggests a co-operative effect of intermediate temperature and cooling rate on the formation of grain boundary serrations.

\subsection{On The Identification of Carbide Species $\&$ Its Role in Serration}
\label{carbide_evolution}

\begin{landscape}
\begin{figure}
  \centering
  \includegraphics[width=1.5\textwidth]{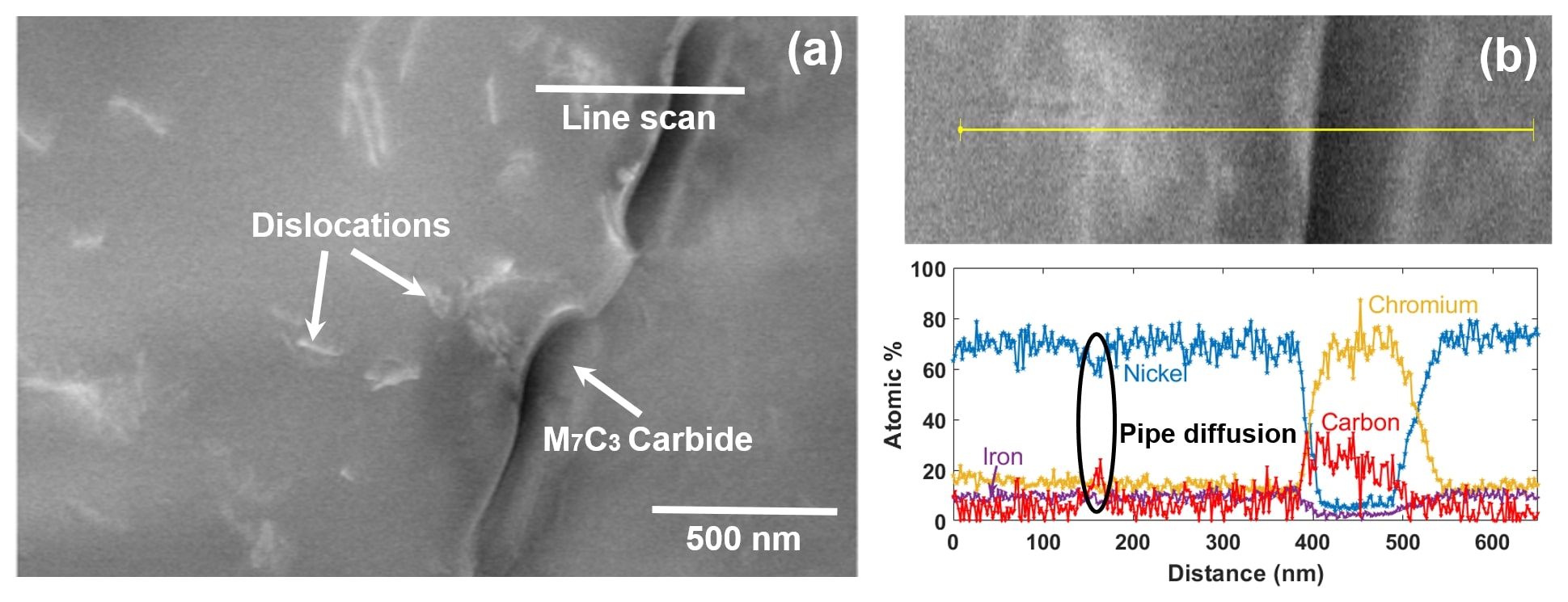}
  \caption{(a): STEM-EDS line scan across the interface of matrix/carbide and (b): the compositional profile acquired.}
  \label{fig:STEM_EDS}
\end{figure}

\end{landscape}

Since carbides clearly play a crucial role in the serration phenomenon in this alloy, further characterisation studies were
carried out. Chemical composition analysis using STEM-EDS is presented in \Fref{fig:STEM_EDS}. Here, a line scan was performed over the matrix/carbide interface. In the carbide, the metal/carbon ratio was around 7:3, suggesting the $\rm M_{7}C_3$ type carbide. A slight decrease of carbon content over distance was also observed, caused in all likelihood by an overlay of carbide and matrix, consistent with a more gradual composition profile. Interestingly, a spike of carbon concentration was captured while scanning through a cluster of dislocations, which is suggestive of pipe diffusion via these crystal defects.

\begin{landscape}
\begin{figure}
  \centering
  \includegraphics[width=1.8\textwidth]{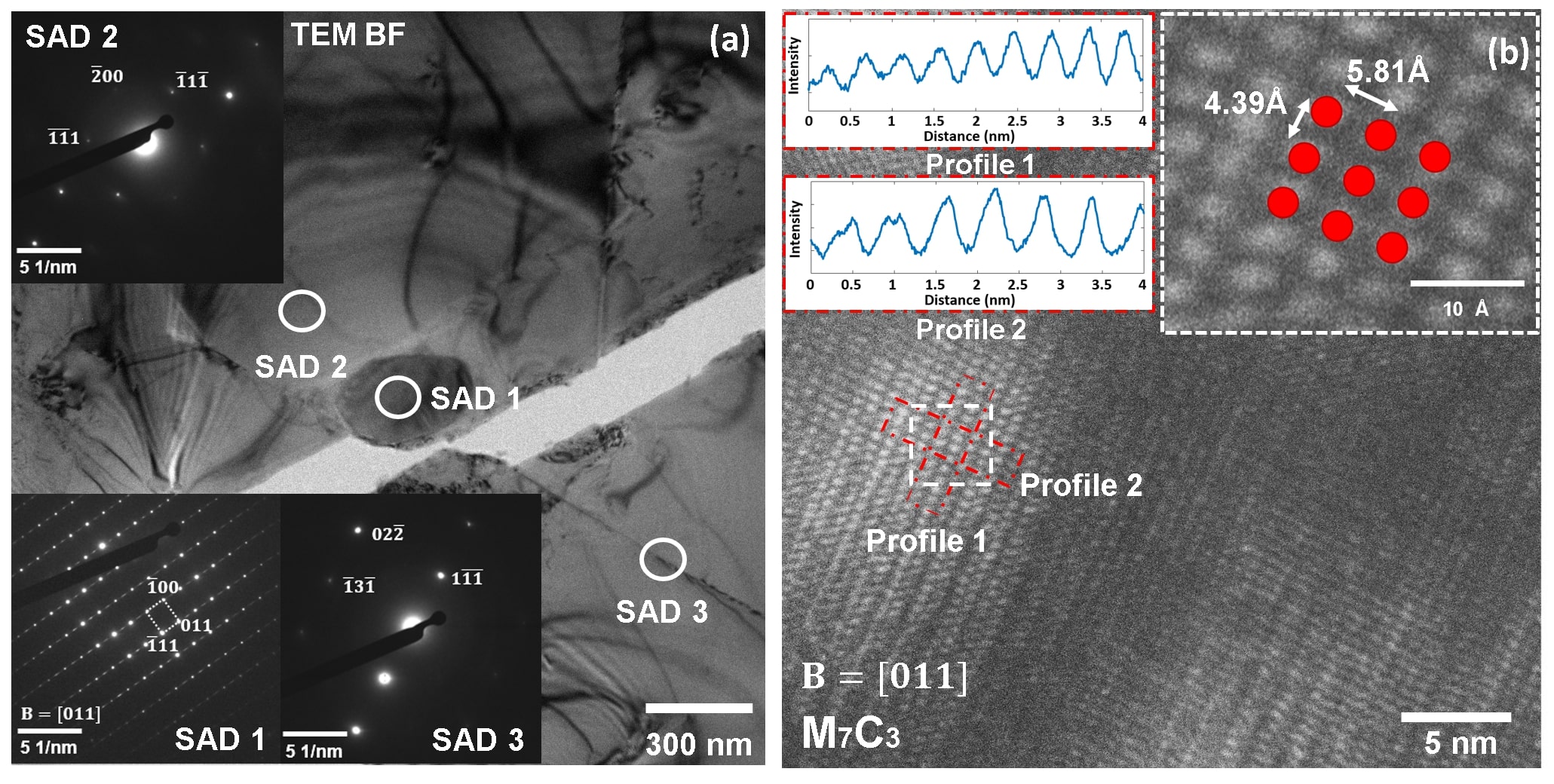}
  \caption{(a): TEM micrograph and selected area diffraction patterns at carbide and both sides of the matrix in the as received material. (b): HR-TEM showing the atomic spacing of the carbide.}
  \label{fig:TEMAR}
\end{figure}

\end{landscape}

\begin{landscape}
\begin{figure}
  \centering
  \includegraphics[width=1.8\textwidth]{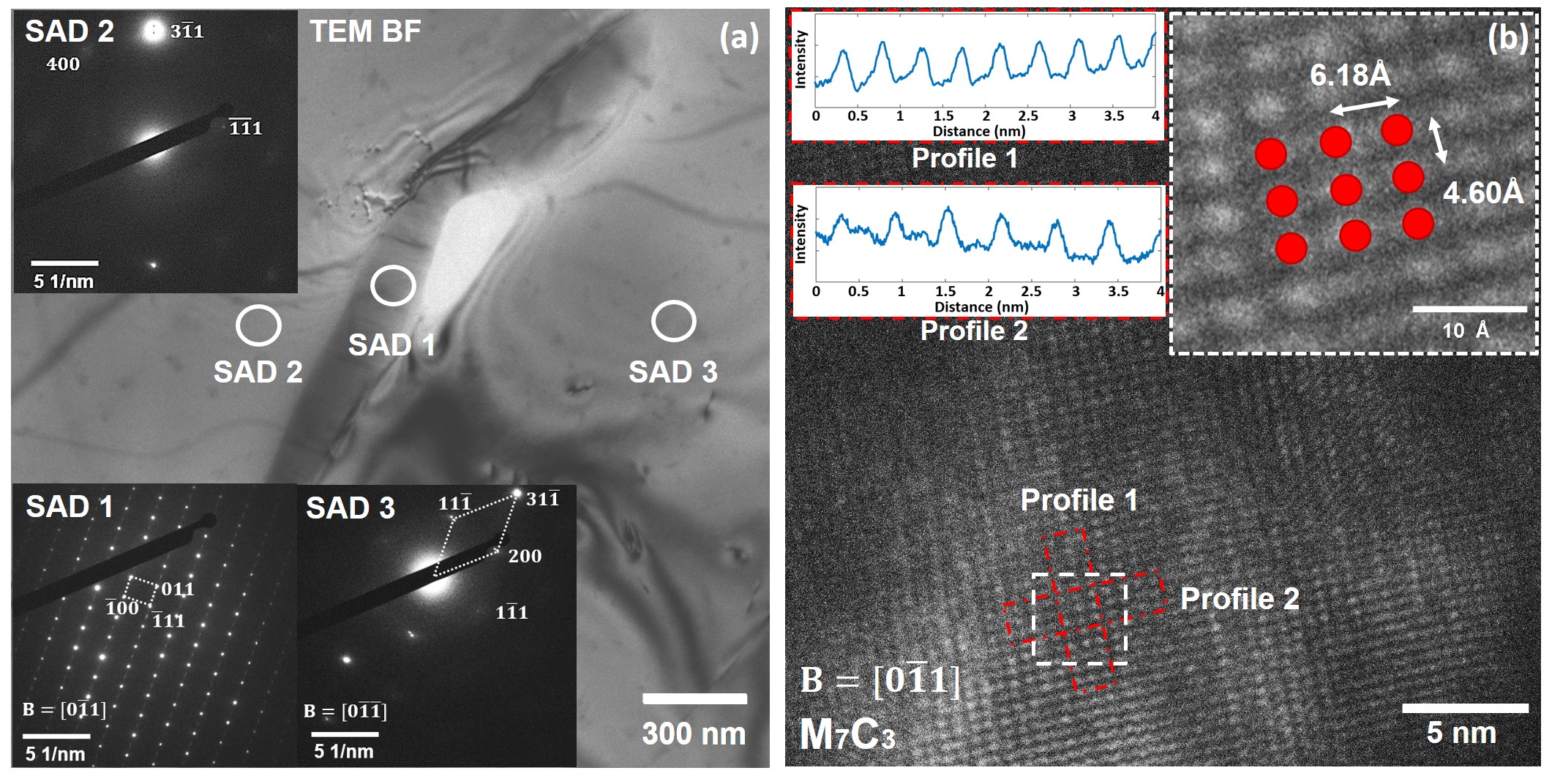}
  \caption{(a): TEM micrograph and selected area diffraction patterns at carbide and both sides of the matrix in the material with serrated grain boundaries. (b): HR-TEM showing the atomic spacing of the carbide.}
  \label{fig:TEMserr}
\end{figure}

\end{landscape}

TEM diffraction experiments were performed on three different zone axes for the determination of the carbide crystal structure in both the as-received and serrated grain boundary specimen (B7). As displayed in \Fref{fig:TEMAR} and \Fref{fig:TEMserr}, two intergranular carbide particles and their surrounding $\gamma$ matrices were also indexed. The carbides anaylsed in both specimens were determined as orthorhombic $\rm M_{7}C_3$ type, where d1 and d2 spacing were found to be consistent with reports in the literature \cite{Villars2006}. Additionally, the plate-like carbide in \Fref{fig:TEMserr} demonstrated a coherent interface with the matrix (right) next to it. The orientation relationship can be expressed by $\{111\}_{\gamma}$//$\{100\}_{\rm M_{7}C_3}$, the interface plane is not determined. Nevertheless, the carbide present in the as-received state was incoherent with both adjacent $\gamma$ matrix grains. HR-TEM performed on both carbides facilitated direct measurements of the d spacing in the $\{100\}_{\rm M_{7}C_3}$ direction, where the lattice constant was measured between 4.39 - 4.60 $\rm \AA$ as opposed to 4.51 - 4.52 $\rm \AA$ by \textit{ab initio} calculation and measurements \cite{Kaplan2013, Villars2006}.

\begin{landscape}
\begin{figure}
  \centering
  \includegraphics[width=1.2\textwidth]{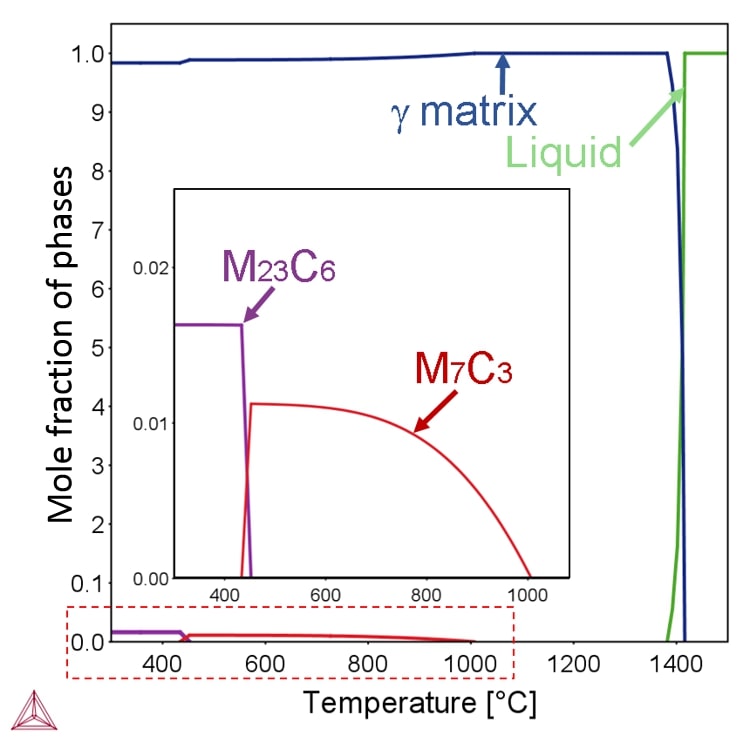}
  \caption{Equilibrium phase diagram of Inconel 600 from 300 - 1500$^\circ$C showing carbide type and dissolution temperature.}
  \label{fig:TC}
\end{figure}
\end{landscape}

At the microscale level, whilst the carbide phase (orthorhombic $\rm M_{7}C_3$) remained unchanged in response to heat treatments, the morphology, shape and distribution became drastically altered. Thorough investigation of the carbides associated with grain boundaries which are serrated resulted in data summarised in Table \ref{carbide}. Numbers for average carbide size, length, thickness, circularity are presented. In addition, carbide nucleation density was calculated by total number of carbide per unit length of grain boundary profile that was measured. To further rationalise the thermodynamic properties of the material studied, a phase diagram was plotted with the ThermoCalc software using the TCNI8 database, as illustrated in \Fref{fig:TC}. The $\rm M_{7}C_3$ carbide dissolution temperature was found at 1007$^\circ$C by calculation, which is close but in error with the experimental value estimated here which is between 1040 - 1070$^\circ$C, as determined by heat treatments B2 $\&$ B12.

\section{Discussion}

The work presented here has shown unambiguously that grain boundary serrations can be induced in the IN600 alloy provided
(i) the solutioning temperature is sufficiently high and (ii) if the cooling rate is slow enough. The serration effect
found here is quite distinct from that reported in the literature for other systems \cite{McQueen1995, Mcqueen2007, Hogg2007, Zouari2016, Gong2016} for which significant deformation is needed, associated with bulging of dislocations cell walls or nucleation of nano-sized grains where dynamic recrystallisation had initiated \cite{Doherty1997, Chen2015}. Here, deformation is not required and the serrations may be defined as \textit{irregularly shaped architecture with saw/wave-like appearance of a random high angle grain boundary that is produced via heat treatment involving an annealing process.} Nonetheless, a detailed consideration of the
formation mechanism for serrations in IN600 is warranted, and this now follows.

\subsection{Formation Mechanisms For Grain Boundary Serration in IN600}

One needs to revisit \Fref{fig:gb_sch} where it was proposed that grain boundaries fall into one of six categories, the last
three of which are associated with serration. \Fref{fig:gb_sch} (d) possessed a grain boundary profile that appears smoother and more wave-like. However, (e) exhibited faceted carbides that grow in a specific orientation, and hence the grain boundary traces appear saw-tooth shaped.

The first mechanism suggested explains the serration type seen in \Fref{fig:gb_sch} (d). Grain boundary migration between pinning particles is the cause, as demonstrated by the bright field images in \Fref{fig:migration_gb}, where the upper grain is attempting to migrate toward the lower one but is impeded by carbides. Migrating boundaries are revealed with more detail in \Fref{fig:migration_gb} (b) and (c). Multiple arrays of dislocation lines are emitted at specific orientations as the grain boundary migrated from the upper grain to the lower one. Clearly, this Zener-type pinning-induced grain boundary serration suggests the necessary grain growth can still occur during slow cooling; however, the pressure from the intergranular carbide counteracts the material's desired boundary motion and inhibits it. Consequently, the part of a boundary in contact with carbide stays immobile while the remaining part moves marginally. For more detailed elucidation of driving force for such situation, we have performed TKD scanning on one of those boundaries. The grain average misorientation (GAM) map in \Fref{fig:TKD_serr} (b) shows the lower grain acquires a higher degree of misorientation as opposed to the upper grain, which favours grain boundary migration as it reduces crystal misorientation. This mechanism is operative whilst carbide nucleation density is not very high: there is capacity for the boundaries to migrate between the carbides.

A second mechanism apparent in \Fref{fig:facet_carbide} is due to the impact of faceted growth of carbides. As frequently reported in the literature \cite{Hong2009, Jiang2012}, carbides such as $\rm M_{23}C_6$ can grow along the habit plane of $\{$1 1 1$\}$ shared by carbide and matrix. It is proposed here that a similar phenomenon is happening. The coherency relationship between the $\rm M_{7}C_3$ carbide and one side of the matrix can be expressed as $\{111\}_{\gamma}$//$\{100\}_{\rm M_{7}C_3}$ in the present case. The $\rm M_{7}C_3$ carbide first nucleates at the grain boundary, and then it follows a preferred growth direction and thus becomes faceted. The specific growth orientation is likely to associate with the coherency stress that is exhibited at the interface for elastic energy benefits. Consider \Fref{fig:facet_carbide} (b): the growth direction is $\sim$30$^\circ$ anticlockwise to the vertical axis, and the carbide has clearly distorted the grain boundary such that a curvature is created from A to B. In addition, due to the curvature introduced by the carbide, an array of dislocations was emitted where the curvature ends at point B. On the other side of the carbide as illustrated in \Fref{fig:facet_carbide} (c), a cluster of dislocations were generated to compensate plasticity induced by both the carbides adjacent to it.

For further evidence of the carbide growth in relation to grain boundary and surrounding matrix, cross-correlation analysis of the TKD patterns were conducted at a more local region of the same carbides. \Fref{fig:Strain_tensor_N} shows the elastic strain tensor calculated from the measured on-axis TKD patterns. For better interpretation, the strain matrix is given varying the X $\&$ Y reference axes defined by the carbide growth direction. Admittedly, the strain state is complex around the carbide, which is believed to be associated with a combination of coherency strain at the interface during carbide growth and the thermal contraction during heat treatment. The difference in linear thermal expansion coefficients reported for $\rm M_{7}C_3$ and $\gamma$ matrix is 4.2 $\times$ 10$^{-6}$K$^{-1}$ \cite{Tuleja2017}, therefore, the thermal strain estimated is on the order of 10$^{-3}$, which is in the same magnitude of measured elastic strain. The coherency misfit, however, is more significant by estimation. The spacing of $\{100\}_{\rm M_{7}C_3}$ plane is larger than $\{111\}_{\gamma}$ spacings by about 10 $\%$ according to the literature \cite{Ernst2011}. The value is significantly large, which suggests the actual interface may be semi-coherent at some regions. The plasticity involved, as reflected by dislocation activities in the vicinity of grain boundary and carbides, is thought to originate from volume changes due to phase transformation from $\gamma$ matrix to carbide. The unit cell volume per metal atom of $\rm Cr_{7}C_3$ is ${\Omega}_{\rm Cr_{7}C_3}$ = 0.01376 nm$^{3}$ according to \cite{Rouault1970}. In contrast, the lattice parameter for FCC $\gamma$ matrix in Inconel 600 is measured at 0.3549 nm \cite{Raju2004}, which gives the atom volume ${\Omega}_{\gamma}$ = (${a_\gamma}^{3}$)/4 = 0.01118 nm$^{3}$. The volume discrepancy between the two is estimated at $\sim$$23\%$ (corresponding to linear strain of  $\sim$$7\%$), which will generate marked compressive stress thus triggering dislocation events. \Fref{fig:Rot_tensor} illustrates the rotation tensor and the geometrically necessary dislocation (GND) density map. A notable amount of lattice rotation near the carbides has induced higher GND density near the grain boundary and around the carbides. Moreover, this is validated by the SEM-TKD BF images in \Fref{fig:facet_carbide}, where a number of dislocation lines were revealed. The locations of dislocation activities \textit{i.e.} discrete, arrays and clusters, are shown to correlate well with the GND density map. It is deduced that the dislocations at the grain boundary have arisen in association with the growth of carbides.

In summary, the sequence of events happening is likely to be the following. During annealing, only limited dislocation content is present and grain boundaries are clear of carbides. When cooling starts, carbides are formed first along the grain boundaries that grow with a preferred orientation. At the same time as carbide growth, local volume changes as Cr $\&$ C atoms form $\rm Cr_{7}C_3$ generate dislocations around the carbide to accommodate plasticity. Grain boundaries are still relatively mobile (over 1000$^\circ$C) at this stage and can migrate in some extent, or the adjacent facetted carbides can twist the boundary while they grow. In both cases, the precipitation and serration are generated simultaneously. Of course the two mechanisms identified above may act together to different extents so that both may be operative at any given boundary. It is probably the case that the precise mechanism operative may be difficult to identify on a two-dimensional section of a
2D SEM/EBSD image because two degrees of freedom is not accounted for. It is proposed that boundaries of the type in
\Fref{fig:gb_sch} (d) are of mixed character.

\subsection{A Process Map and its Limitations}

In this section, a process map is constructed on the basis of the quantitative information acquired for the effect of
solutioning temperature and cooling rate on degree of serration. Clearly, a solution temperature at $1070^\circ$C/120min (B2) has been shown to be insufficient to generate serration during subsequent cooling: the driving force of the serration formation requires the synergy of carbide reprecipitation and grain boundary of sufficient mobility, hence this explains why no serration can be triggered for a solution temperature equal or lower than this. Thus the serration index at $1070^\circ$C equals 0 at all cooling rates by definition, as demonstrated in \Fref{fig:Index_contour} (a). Similarly, A1 is then used as another bounding condition, since a cooling rate around $600^\circ$C/min or higher will not trigger serration, due to the kinetics for re-precipitation on the grain boundary limiting the size of carbides present. The rest of the conditions tested in the matrix are plotted on \Fref{fig:Index_contour} (a), with its measured serration index next to it. The data have been transformed into a processing map by interpolating serration index values onto a contour plot using MATLAB, see \Fref{fig:Index_contour} (b). The serration index is displayed as a function of solution temperature (fixed at 120 mins) and cooling rates with a fixed intermediate temperature at $900^\circ$C.

The processing map confirms the relative influence of cooling rate and solutioning temperature, as illustrated in \Fref{fig:correlation}, and summarises the situation for IN600. For any particular cooling rate, solutioning temperature has a significant influence on the degree of serration, while with decreasing solution temperature the serration is then suppressed. The cooling rate also plays a crucial role in the formation of serrations --- with a solution temperature at $1140^\circ$C, the degree of serration declines drastically from 0.25 - $60^\circ$C/min, in agreement with other reports \cite{Hong2009, Mitchell2009}. A logarithmic relationship between serration index and cooling rate is established, as displayed in \Fref{fig:correlation} (c). \Fref{fig:correlation} (b) presents another logarithmic relationships between carbide nucleation density and cooling rate, consistent with the serration index being a function of carbide nucleation frequency. \Fref{fig:correlation} (d) shows an inverse linear relation between serration index and nucleation density generated as fewer particles are nucleating on a grain boundary.

The data of \Fref{fig:Index_contour} contains one anomaly which is worthy of consideration. For a solution treatment of
$1100^\circ$C --- from the above analysis and deduction --- one would expect to see a higher degree of serration with very slow cooling rate, yet observations made under these conditions do not confirm this, see \Fref{fig:gb_corner}. A plausible
explanation may be as follows. Since the heat treatment of $1100^\circ$C/120 mins dissolves the carbide only partially and increases the grain size rather moderately, this may lead to fewer available carbon atoms having the necessary potency for precipitation in an even larger grain boundary area. As a result, carbide nucleation density may be lower compared to higher solutioning temperatures. Therefore, the precipitation may become more selective with a preference for grain triple junctions. Some carbides show a tendency to facet or impede grain boundary migration in \Fref{fig:gb_corner}, however, the degree of serration is negligible overall due to carbides being far apart from each other resulting in only a moderate effect. This situation contrasts with that at faster cooling rate: for example $12 ^\circ$C/min, when a larger undercooling in comparison leads to more carbides being formed on the grain boundary and thus more frequent local faceting/migration.

Finally, a stepped isothermal heat treatment has been implemented for determining the necessity for slow cooling in regard to serration formation. One may consider using repeated isothermal holding to mimic the slow cooling process, see \Fref{fig:step_age}. In comparison to B7, both treatments had $1140^\circ$C/120 minutes solution treatment and intermediate temperature of $900^\circ$C/120 minutes. Instead of obtaining a continuous cooling process, the new heat treatment acquired nine segments of isothermal hold followed by quenching, in an increment of $30^\circ$C for 120 minutes. In contrast to thick planar intergranular carbides as shown in \Fref{fig:step_age} (c), the step aged sample obtained both inter/intragranular carbides. The continuous granular carbides formed on the grain boundaries with a high nucleation density as a result of quenching, and then coarsened with subsequent isothermal holding. This
further experimental test confirms that slow cooling from a solution treatment passing through carbide dissolution temperature is 
an essential condition for the serration effect, and serves as further evidence that this effect is now better understood.

\section{Summary and Conclusions}

The phenomenon of grain boundary serration in Inconel 600 has been studied. The following specific conclusions can be drawn from
this work:

\begin{enumerate}

\item{A new approach has been established for quantification of grain boundary serration based on analysis by Fast Fourier Transformation of the boundary trace. The method utilises a combination of electron microscopy, image analysis and mathematical transformation to analyse the serrated grain boundary morphologies. This represents a new quantitative approach to serration assessment.}

\item{The basis of the method involves the wavelength and amplitude of the three strongest sinusoids being combined into one metric that allows the degree of serration to be described. A physical interpretation of the serration index is the area subtended by the undulating grain boundary on a plane, polished section.}

\item{The conditions needed to trigger serration during annealing have been identified. It is confirmed unambiguously that solution treatment followed by slow cooling is needed; isothermal holding is not sufficient to cause this effect. The presence of carbide locally on the grain boundary does not appear to be a sufficient condition for serration to be initiated.}

\item{Carbides present in IN600 have been identified as orthorhombic $\rm M_{7}C_3$ type in both the as-received state and the heat-treated state. Carbides showed a coherency relationship to one side of the matrix after heat treatment, where the orientation relation can be expressed by $\{111\}_{\gamma}$//$\{100\}_{\rm M_{7}C_3}$.}

\item{Two serration formation mechanisms are proposed. A first involves sections of the grain boundary being pinned by intergranular carbides while other sections migrate through gaps between the carbides. A second mechanism requires faceted carbide growth, which eventually distorts the boundary. In some cases, both mechanisms have been observed to operate simultaneously. Which of the two formation mechanism occurs is dominant depends upon the solutioning temperature and nucleation density of the carbide precipitates.}

\item{The new serration index and analysis carried out has allowed a processing map for serration to be proposed.}

\end{enumerate}

\section{Acknowledgments}

Yuanbo Tang acknowledges financial support from St Edmund Hall, University of Oxford through William R Miller Award. Tang also shows gratitude to Dr Yilun Gong and Dr Shaokai Zheng for inspiring discussions.

\newpage
\clearpage
\section{References}

\bibliography{tangrefsall}

\newpage
\clearpage

\section{Tables and Figures}

\begin{table}[h]
\centering
\caption{Alloy composition for current study in wt-$\%$ (Ni-base), measured using ICP-OES, ICP-combustion (Carbon) and Spark OES (Boron)}
\label{table: composition}
\begin{tabular}{l*{10}{c}r}
\hline \hline
Name & Ni       & Cr       & Fe      & C          & Co      & B      & Ti  \\
\hline
Inconel 600      & Bal     & 15.75     & 9.12        & 0.071   & 0.04     & 0.004     & 0.22\\
\hline \hline
\end{tabular}
\end{table}

\begin{landscape}

\begin{table}[h]
\centering
\caption{Heat treatment matrix that performed in current study. First letter in label refers to heat treatment type described in \Fref{fig:HT}.}
\label{HT_matrix}
\begin{tabular}{l*{10}{c}r}
\hline \hline
  & Solution treatment & Cooling & Intermediate treatment  & Cooling & Ageing  & Serration & Boundary type  \\
\hline
Unit        &   $^\circ$C   & $^\circ$C/min     & $^\circ$C  & $^\circ$C/min  & $^\circ$C $\&$ min   &   N/A  & N/A  \\
\hline \hline
A1 &  1140/2hrs & 600  & 22     & N/A      & N/A     & N & a \\
\hline
A2 & 1140/2hrs & WQ  & 22     & N/A      & N/A     & N & a   \\
\hline \hline
B1 &  1000/2hrs & 12  & 900     & WQ      & N/A     &N &  c   \\
\hline
B2 &  1070/2hrs & 12 & 900     & N/A      & N/A     & N &  c \\
\hline
B3 &  1100/2hrs & 0.25  & 900  & WQ      & N/A     & N & c \\
\hline
B4 &  1100/2hrs & 12  & 900   & WQ      & N/A     & Y  & d - f \\
\hline
B5 &  1100/2hrs & 30  & 900   & WQ      & N/A    & Y &  d - f\\
\hline
B6 &  1120/2hrs & 3  & 900     & WQ      & N/A     & Y & d - f \\
\hline
B7 &  1140/2hrs & 0.25  & 900  & WQ      & N/A     & Y & d - f \\
\hline
B8 & 1140/2hrs & 3  & 900    & WQ      & N/A     & Y & d - f \\
\hline
B9 &  1140/2hrs & 12  & 900     & WQ      & N/A     & Y & d - f \\
\hline
B10 &  1140/2hrs & 30  & 900     & WQ      & N/A     & Y & d - f \\
\hline
B11 &  1140/2hrs & 60  & 900     & WQ      & N/A     & Y & d - f \\
\hline
B12 &  1140/2hrs & 0.25  & 1040     & WQ      & N/A     & Y & d - f \\
\hline \hline
C1 &  1140/2hrs & 0.25  & 1040/30min     & WQ      & N/A     & Y & d - f \\
\hline
C2 &  1140/2hrs & 12  & 900/30min     & WQ      & N/A     & Y & d - f \\
\hline
C3 &  1140/2hrs & 12  & 1040/30min   & WQ      & N/A     & N & a \\
\hline
C4 &  1140/2hrs & 12  & 1060/30min    & WQ      & N/A     & N & a \\
\hline \hline
D1 &  1140/2hrs & 12  & 1040/30min     & WQ      & 900/30min     & N & b \\

\hline \hline
\end{tabular}
\end{table}

\end{landscape}

\newpage
\clearpage
\begin{landscape}

\begin{table}[h]
\centering
\caption{Carbide and serration quantification matrix. }
\label{carbide}
\begin{tabular}{l*{10}{c}r}
\hline \hline
&\multicolumn{2}{c}{Heat Treatment} & \multicolumn{5}{c}{Carbide} & \multicolumn{4}{c}{Serration} \\
\hline
 & Solution & Cooling rate  & Size & Length & Thickness& Circularity & Nucleation density & Wavelength & Amplitude &  Index &  Std Dev\\
\hline
 & $^\circ$C &  $^\circ$C/min     & $\mu$m$^2$     &   $\mu$m   & $\mu$m & N/A  & mm$^{-1}$ &$\mu$m &$\mu$m & $\mu$m$^2$ & $\mu$m$^2$ \\
\hline \hline

B4 &   1100/2hr & 12  &   0.35 & 2.05 & 0.37 & 0.33 & 276.3 & 19.02 & 0.64 & 12.07 & 7.39\\
\hline
B5 &   1100/2hr & 30 &   0.25 & 1.19 & 0.34 & 0.49 & 326.3 & 13.70 & 0.62 & 8.47 & 3.02\\
\hline
B6 &   1120/2hr & 3  &   0.56 & 2.40 & 0.55 & 0.41 & 288.0 & 21.60 & 0.75 & 16.12 & 9.30\\
\hline
B7 &   1140/2hr & 0.25  &   6.36 & 6.99 & 1.72 & 0.36 & 109.2 & 24.54 & 0.92 & 22.65 & 15.10\\
\hline
B8 &   1140/2hr & 3  &   0.67 & 2.06 & 0.49 & 0.38 & 371.0 & 23.80 & 0.77 & 18.22 & 10.80 \\
\hline
B9 &   1140/2hr & 12  &   0.34 & 1.71 & 0.37 & 0.43 & 417.3 & 23.44 & 0.58 & 13.62 & 8.26\\
\hline
B11 &   1140/2hr & 60  &   0.09 & 0.59 & 0.22 & 0.70 & 694.6 & 21.83 & 0.47 & 10.35 & 2.91\\
\hline
C1 &   1140/2hr & 0.25  &   2.45 & 5.78 & 1.00 & 0.27 & 116.2 & 21.03 & 0.59 & 12.41 & 8.35\\
\hline
C2 &   1140/2hr & 12 &  0.44 & 1.86 & 0.44 & 0.45 & 432.1 & 26.15 & 0.62 & 16.15 & 10.69\\
\hline

\hline \hline
\end{tabular}
\end{table}

\end{landscape}

\newpage
\clearpage

\newpage
\clearpage

\newpage
\clearpage

\newpage
\clearpage

\newpage
\clearpage

\newpage
\clearpage

\newpage
\clearpage

\newpage
\clearpage

\newpage
\clearpage

\newpage
\clearpage

\begin{landscape}

\newpage
\clearpage

\begin{figure}
  \centering
  \includegraphics[width=1.5\textwidth]{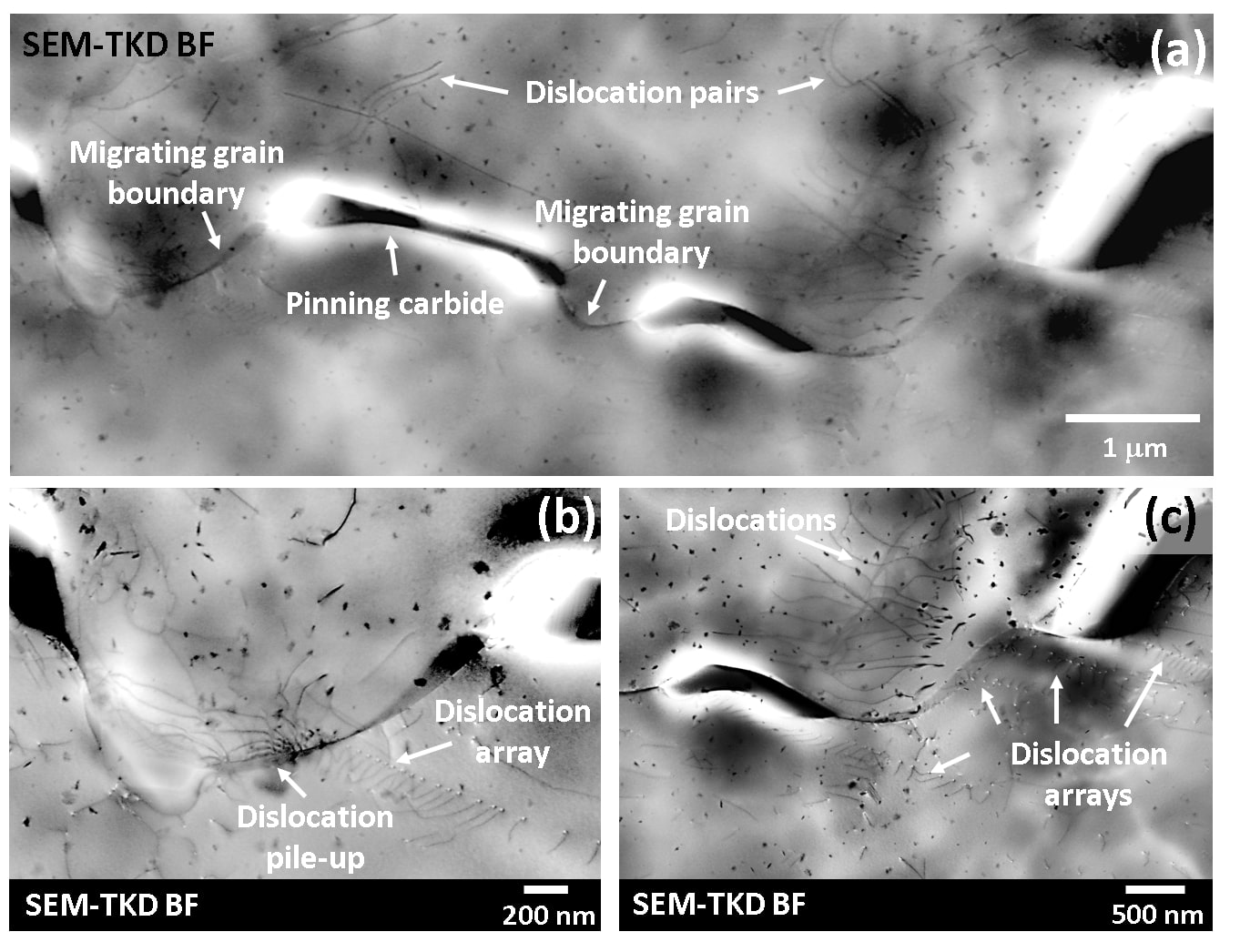}
  \caption{Grain boundary deformation due to migration pinned by carbides (a) and local features in higher magnifications (b) $\&$ (c).}
  \label{fig:migration_gb}
\end{figure}

\begin{figure}
  \centering
  \includegraphics[width=1.5\textwidth]{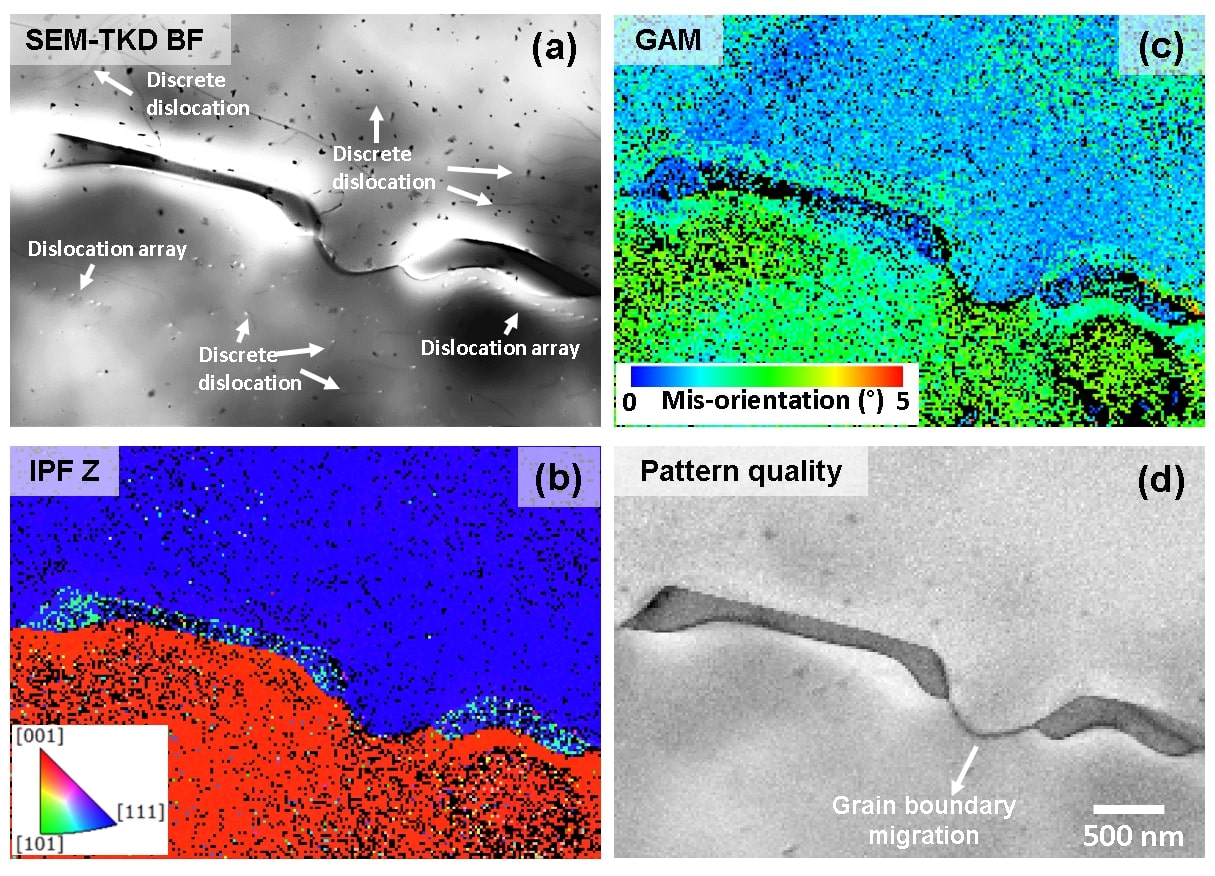}
  \caption{Grain boundary migration impeded by 2 carbides leads to serration. (a): Bright Field (BF) Argus image showing dislocation in arrays and discretely. (b-d) Grain average misorientation (GAM) map, inverse pole figure (IPF) from Z direction and pattern quality map.}
  \label{fig:TKD_serr}
\end{figure}

\begin{figure}
  \centering
  \includegraphics[width=1.5\textwidth]{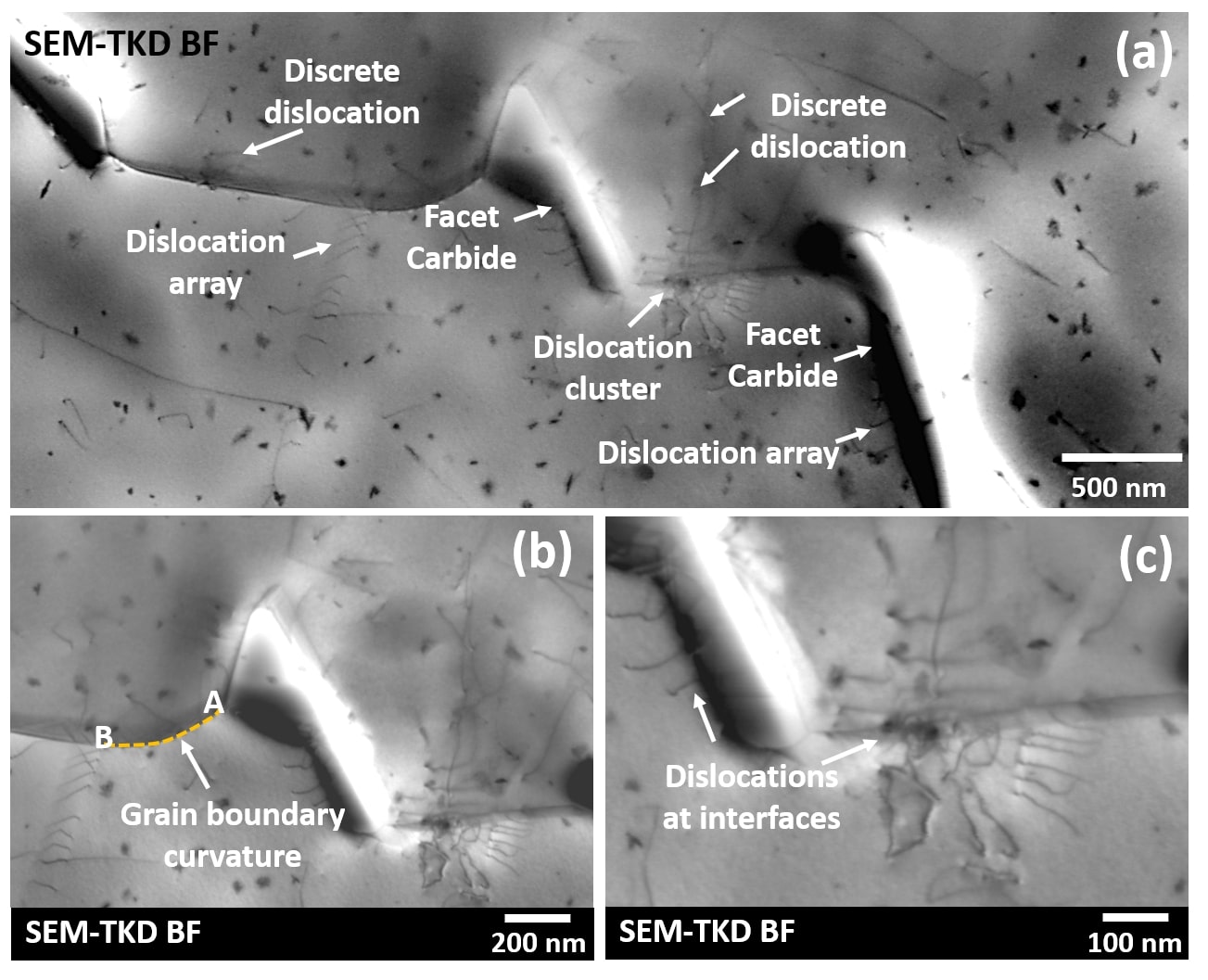}
  \caption{Grain boundary distorted by facet carbides (a) and local features in higher magnifications (b) $\&$ (c).}
  \label{fig:facet_carbide}
\end{figure}

\begin{figure}
  \centering
  \includegraphics[width=1.5\textwidth]{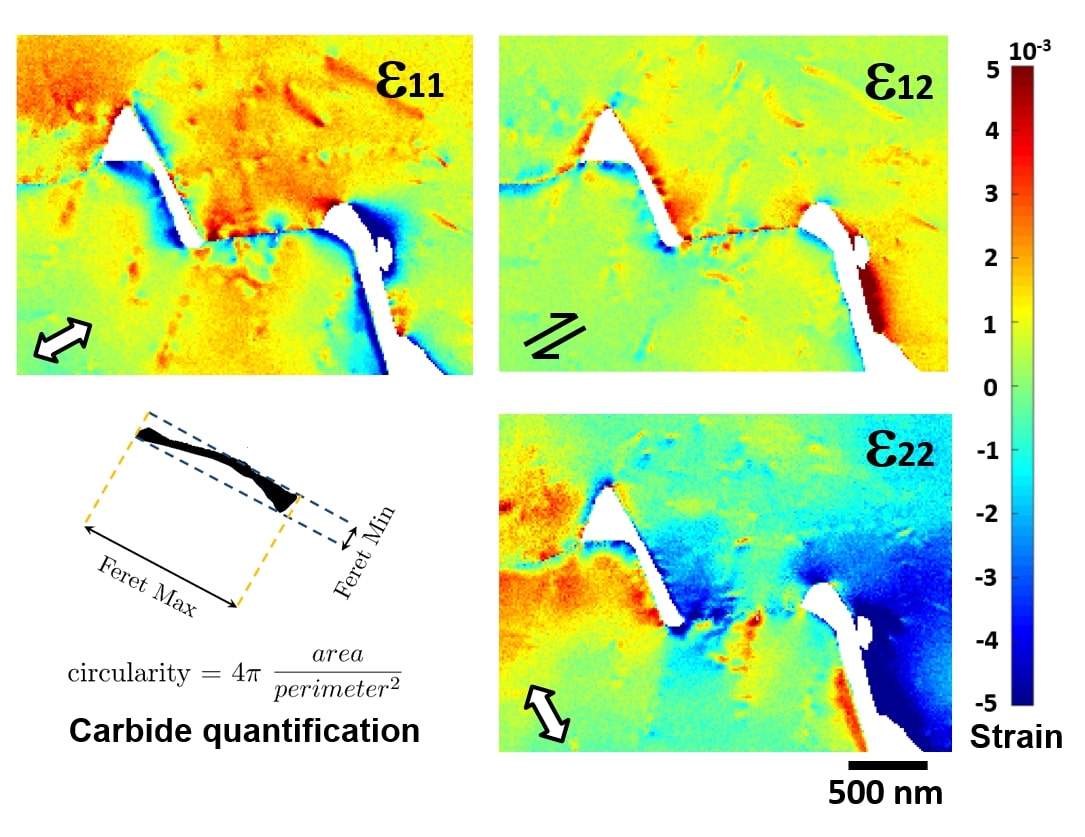}
  \caption{Elastic strain tensor map for two faceted carbides on a grain boundary exhibiting coherency strain induced serration and description of carbide measurements using Feret max/min.}
  \label{fig:Strain_tensor_N}
\end{figure}

\begin{figure}
  \centering
  \includegraphics[width=1.5\textwidth]{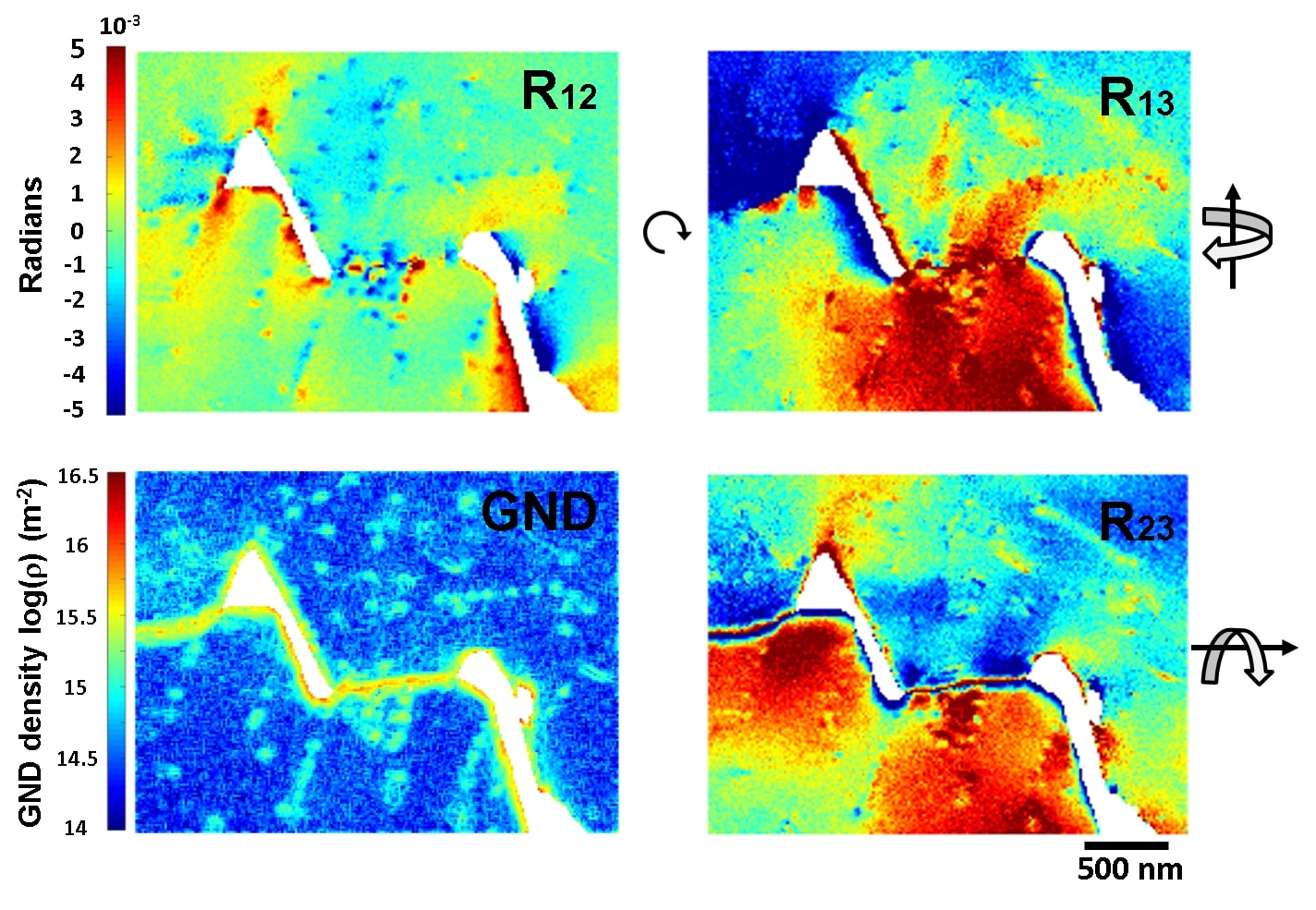}
  \caption{Rotation tensor and geometrically necessary dislocation (GND) density map for two faceted carbides on a serrated grain boundary.}
  \label{fig:Rot_tensor}
\end{figure}

\end{landscape}

\newpage
\clearpage
\begin{landscape}
\begin{figure}
  \centering
  \includegraphics[width=1.5\textwidth]{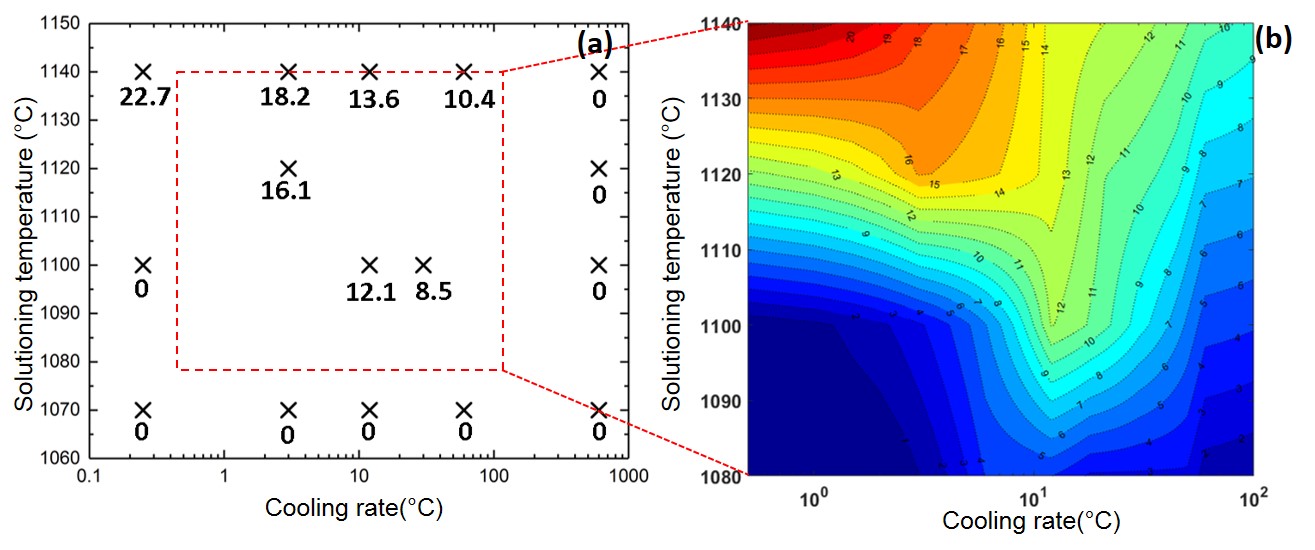}
  \caption{(a) Serration index obtained for a range of solution temperature and cooling rate. Crosses indicate actual data points and the number below them are serration index measurements. (b) Contour plot of serration index as a function of solution temperature and cooling rate. Solution treatment time is 120 minutes and intermediate temperature is 900$^\circ$C in both graphs.}
  \label{fig:Index_contour}
\end{figure}

\begin{figure}
  \centering
  \includegraphics[width=1.2\textwidth]{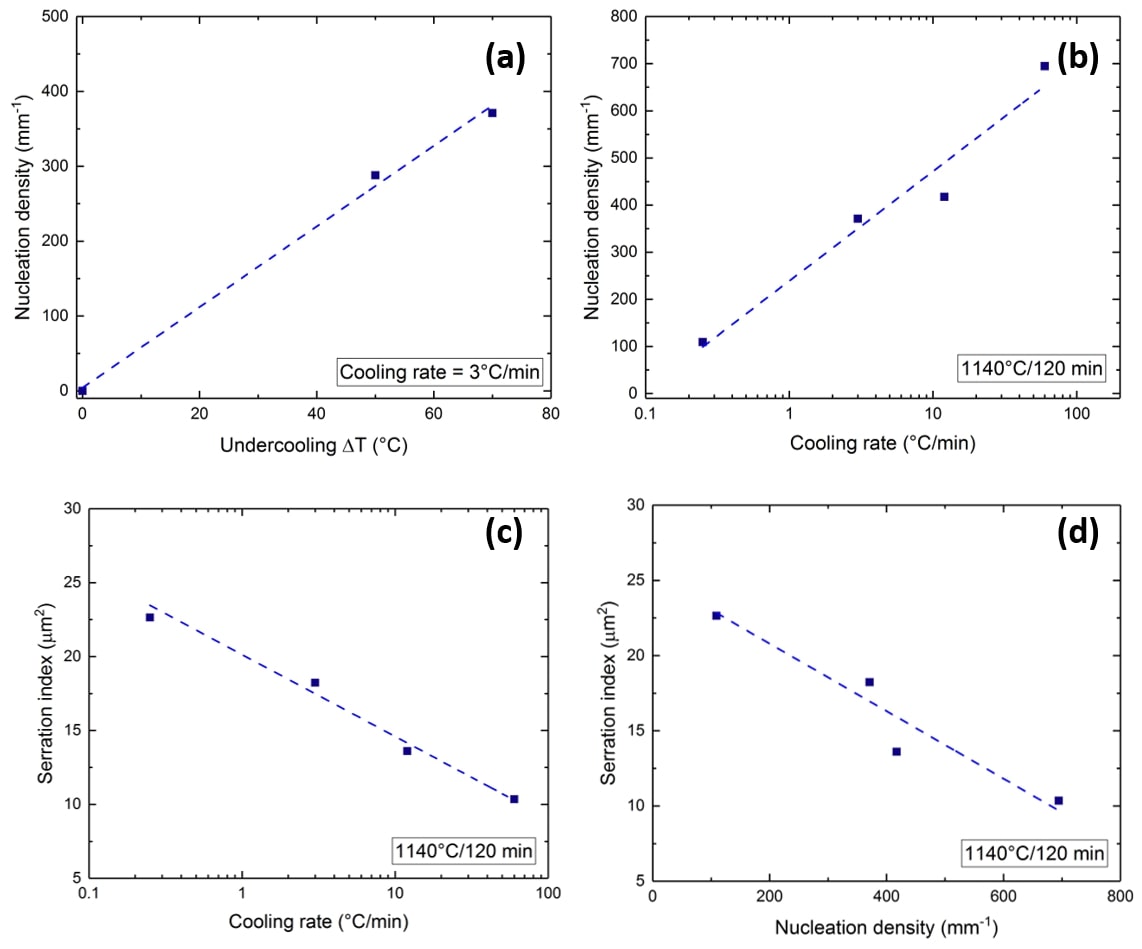}
  \caption{Relationship between processing parameters, carbide distribution and serration index. (a): Nucleation density in relation to the amount of undercooling at 3$^\circ$C/min. (b): Nucleation density is a logarithmic function of cooling rate. (c): Serration index as a logarithmic function of cooling rate and (d): Serration index in adverse relation to nucleation density of carbides. Scatter points are actual data points and dotted lines are linear best fit lines.}
  \label{fig:correlation}
\end{figure}

\newpage
\clearpage
\begin{figure}
  \centering
  \includegraphics[width=1.5\textwidth]{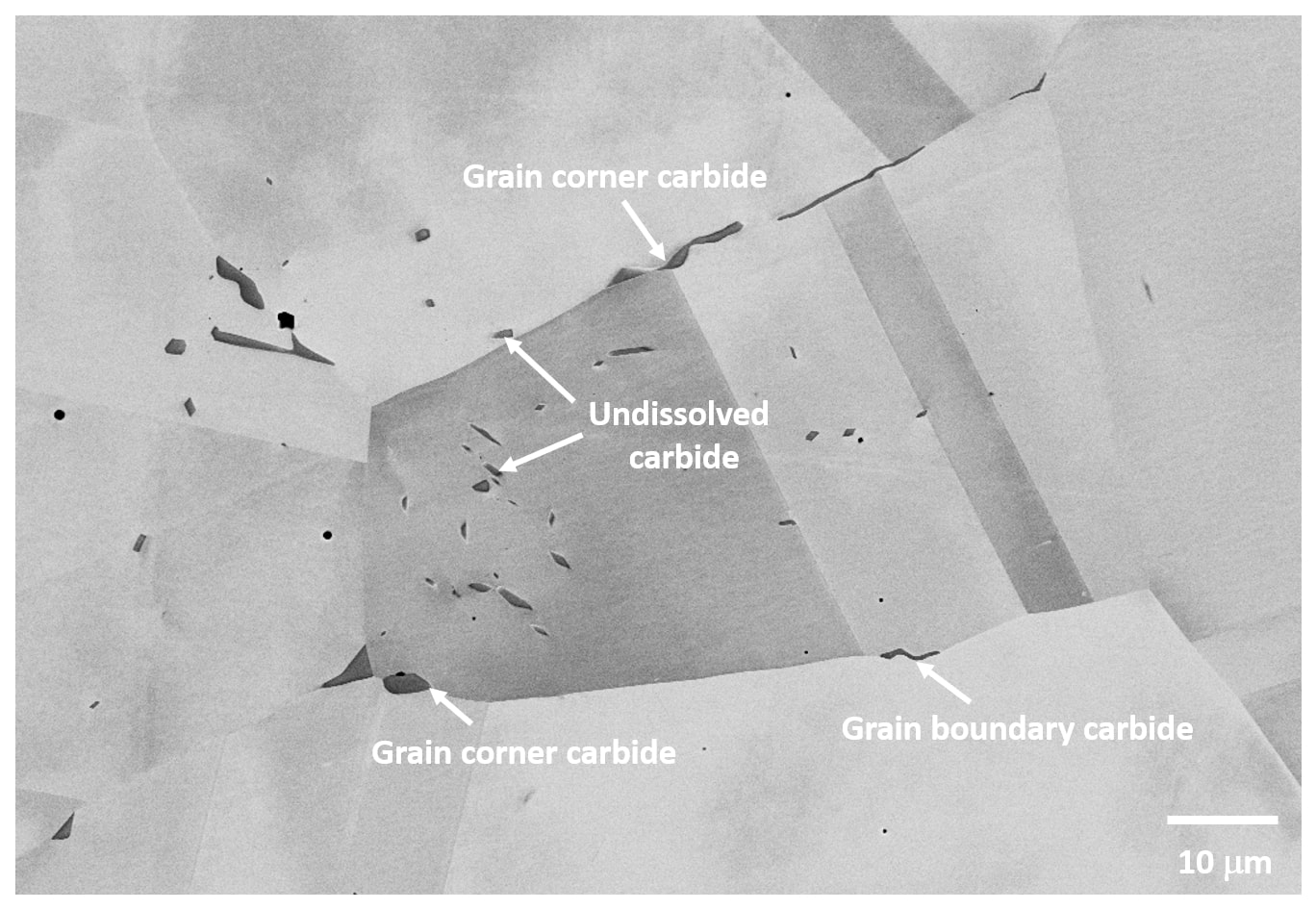}
  \caption{Microstructure of B3 that obtains non-serrated grain boundary. The carbide achieved partial dissolution and remained in grain interior. Reprecipitation of carbides are found primaryly on grain corners and boundaries, which possesses granular and/or blocky morphology.}
  \label{fig:gb_corner}
\end{figure}

\newpage
\clearpage

\begin{figure}
  \centering
  \includegraphics[width=1.5\textwidth]{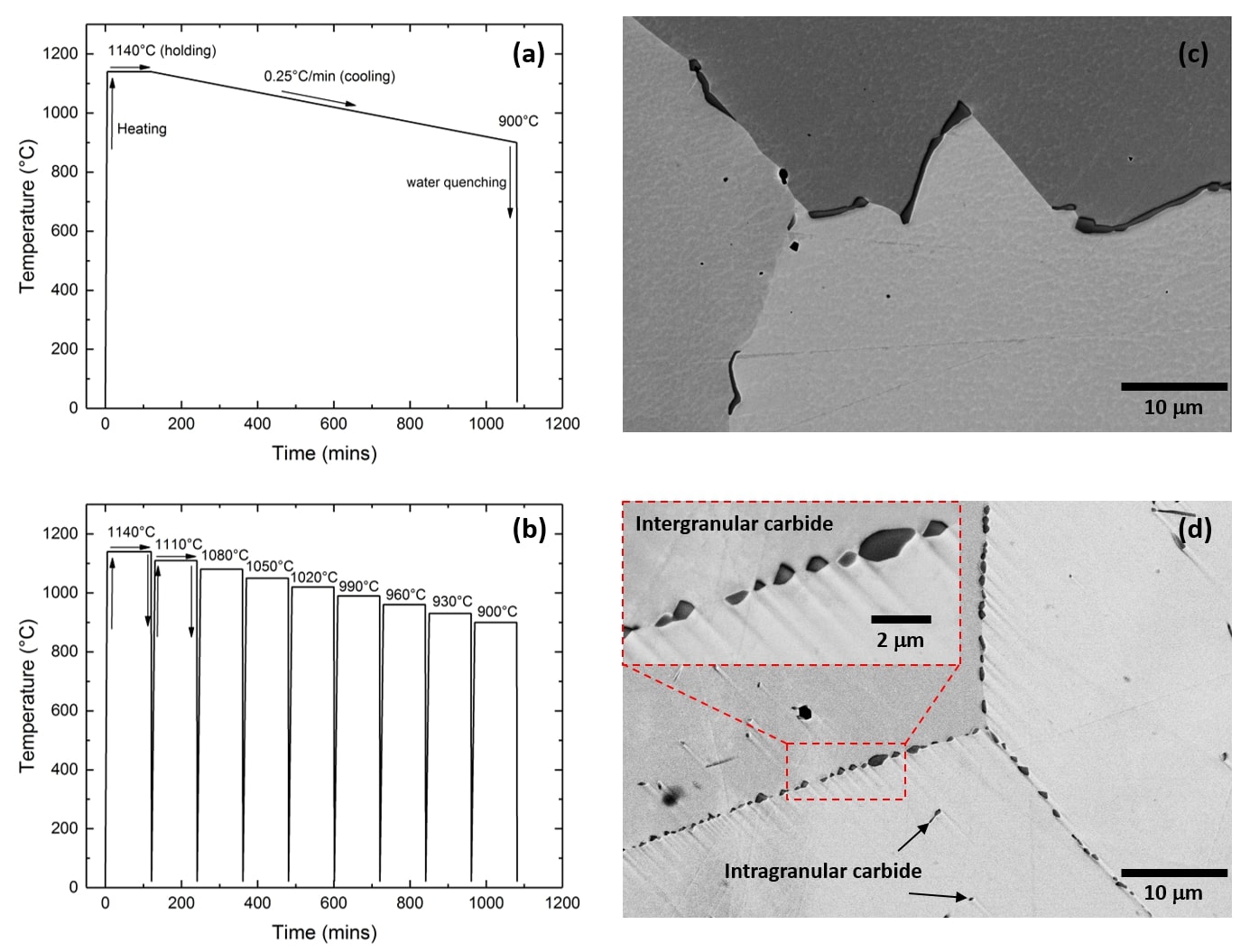}
  \caption{Heat treatment profile for B7 with a step isothermal holding that mimics it, and their corresponding microstructures.}
  \label{fig:step_age}
\end{figure}
\end{landscape}

\end{document}